\documentclass[%
  superscriptaddress,
  twocolumn,
  floatfix,
  10pt,
  aps,
  prl,
  final,letterpaper
]{revtex4-2}

\usepackage{graphicx}
\usepackage{hyperref}
\usepackage[utf8]{inputenc}

\usepackage{amsmath,amsfonts,amssymb}
\usepackage{physics}

\newcommand{\pdegout}[1] {\hat{p}_{#1}^{\text{out}}}
\newcommand{\pdegin}[1] {\hat{p}_{#1}^{\text{in}}}
\newcommand{\pdeginout}[2] {\hat{p}_{#1, #2}^{\text{in,out}}}
\newcommand{\expqdegout} {\hat{q}^{\text{out}}}
\newcommand{\qdegout}[1] {\hat{q}_{#1}^{\text{out}}}

\newcommand{\ulqdegval}[1] {#1}

\newcommand{\ulqdeg}[1] {q_{#1}}

\newcommand{\surv}[1] {P(#1)}
\newcommand{\usurv} {P_\infty}

\begin{document}

\title{Directed Percolation in Temporal Networks}

\author{Arash Badie-Modiri}
\affiliation{Department of Computer Science, School of Science, Aalto University, FI-0007, Finland}
\author{Abbas K.~Rizi}
\affiliation{Department of Computer Science, School of Science, Aalto University, FI-0007, Finland}
\author{Márton Karsai}
\affiliation{Department of Network and Data Science
Central European University, 1100 Vienna, Austria}
\affiliation{Alfr\'ed R\'enyi Institute of Mathematics, 1053 Budapest, Hungary}
\author{Mikko Kivelä}
\affiliation{Department of Computer Science, School of Science, Aalto University, FI-0007, Finland} 

\date{\today}

\begin{abstract}
Connectivity and reachability on temporal networks, which can describe the spreading of a disease, decimation of information or the accessibility of a public transport system over time, have been among the main contemporary areas of study in complex systems for the last decade. However, while isotropic percolation theory successfully describes connectivity in static networks, a similar description has not been yet developed for temporal networks. Here address this problem and formalize a mapping of the concept of temporal network reachability to percolation theory. We show that the limited-waiting-time reachability, a generic notion of constrained connectivity in temporal networks, displays directed percolation phase transition in connectivity. Consequently, the critical percolation properties of spreading processes on temporal networks can be estimated by a set of known exponents characterising the directed percolation universality class. This result is robust across a diverse set of temporal network models with different temporal and topological heterogeneities, while by using our methodology we uncover similar reachability phase transitions in real temporal networks too. These findings open up an avenue to apply theory, concepts and methodology from the well-developed directed percolation literature to temporal networks.
\end{abstract}

\maketitle

Many dynamical processes evolving on networks are related to the problem of reachability. Reachability describes the existence of a possible path of connections between two nodes, denoting the possibility and the extent that one node can affect, cause a change or communicate to the others based on interactions represented in the network. The conception and formalism of reachability, however, changes dramatically if one considers the time-varying nature of connections between nodes~\cite{holme2019temporal} as opposed to the classic static network modeling of systems where connections are considered constant. Time induces an inherent direction of connectivity, as it restricts the direction of influence or information flow. This in turn has an impact on many dynamical processes evolving on such networks, such as spreading \cite{lambiotte2016guide, holme2012temporal, holme2015modern}, social contagion \cite{daley1964epidemics, castellano2009statistical} ad-hoc message passing by mobile agents \cite{tripp2016special} or routing dynamics \cite{nassir2016utility}. In these processes, interacting entities may have limited memory, thereby only building up paths constrained by limited waiting-times, further restricting the eligible temporal structure for their global emergence.

Directed percolation (DP) is a paradigmatic example to characterize connectivity in temporal systems. This process exhibits dynamical phase transitions into absorbing states with a well-defined set of universal critical exponents \cite{hinrichsen2000non, odor2004universality, hinrichsen2006non, henkel2008non}. Since its introduction \cite{broadbent1957percolation} and during its further development \cite{blease1977directed}, directed percolation attracted considerable attention in the literature. It has applications in reaction-diffusion systems \cite{schlogl1972chemical}, star formation in galaxies \cite{gerola1980theory}, conduction in strong electric fields in semiconductors \cite{van1981hopping}, and biological evolution \cite{bak1993punctuated}. While it is straightforward to define idealized models governed by directed percolation, such as lattice models \cite{domany1984equivalence, kinzel1985phase, harris1974contact, jensen1993critical, mendes1994generalized, ziff1986kinetic, dhar1987collapse}, its features are more difficult to realize in nature \cite{hinrichsen2000possible,henkel2008non}, allowing only a few recent experimental realizations of directed percolation \cite{takeuchi2007directed,lemoult2016directed, sano2016universal}. Nevertheless, this description is advantageous in providing an understanding of the connectivity of temporal structures to describe ongoing dynamical processes \cite{barrat2008dynamical, pastor2015epidemic, kempe2002connectivity, moody2002importance, holme2005network, pan2011path, scholtes2014causality, stehle2011high, dai2020temporal, aleta2020data, parshani2010dynamic}.

There is a thorough theoretical understanding of static network connectivity with several concepts borrowed from percolation theory, such as phase transitions, giant components and susceptibility. These concepts, originally developed for lattices and random networks, are routinely used to analyze real-world networks and processes, e.g., disease spreading \cite{newman2002spread, kenah2007second, kenah2011epidemic,rizi2021epidemic, hiraoka2021herd}. Connectivity is also a central property of temporal networks, with several recent techniques to characterize it, e.g., using limited-waiting-time reachability \cite{crescenzi2019approximating,badiemodiri2020efficient,casteigts2019computational,thejaswi2020restless,himmel2019efficient}.

A mapping between temporal reachability phase transition and directed percolation has been anticipated before. This is a straightforward intuition as directed percolation accounts for the time-induced inherent directionality that characterizes temporal networks. For the special cases of contact (SIS) and SIRS processes, this mapping has been shown over a regular lattice structure with the assumption that the contact between nodes follows a Poisson point process \cite{de2010stochastic,henkel2008non,hinrichsen2000non}. This mapping has been shown for a particular class of temporal dynamical systems, involving deterministic walks and discrete temporal layers \cite{parshani2010dynamic}. For a more general class of temporal networks Ref.~\cite{kivela2018mapping} conjectured the mapping with directed percolation based on semantic similarities between the two systems and some empirical evidence. However, these studies could not provide conclusive evidence for this mapping for a broader set of temporal networks. In this work, we aim to show analytically that limited-waiting-time reachability on temporal networks, under a mean-field assumption of connectivity, has a phase transition in the directed percolation universality class. Combined with the experimental results of Ref.~\cite{longpaper}, we conclude that the same is true for a diverse subset of temporal networks, with a wider range of temporal and spatial connectivity compared to the mean-field assumption. Lastly, we illustrate how the directed percolation methodology, formalism and the introduced characteristic quantities can be used to analyze real-world temporal networks, for example, in detecting the onset of reachability phase transitions.

\begin{figure}
    \centering
    \includegraphics[width=1\linewidth]{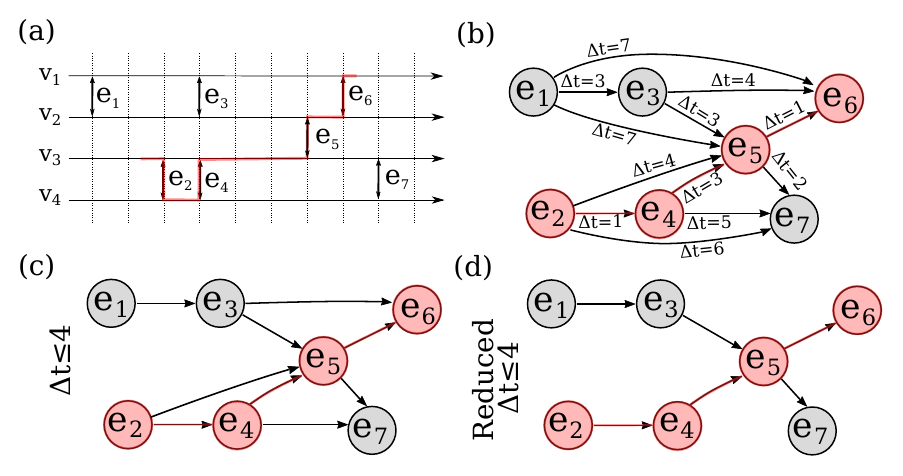}
    \caption{Different representations of an instantaneous, undirected temporal network. (a) Vertices $v_i$ are connected via dyadic instantaneous events $e_j$. (b)
   In a weighted temporal event graph, adjacent events are connected via links directed by time and weighted with the time difference $\Delta t$ between them. Paths in an event graph are equivalent to time-respecting paths \cite{saramaki2019weighted}. (c) Waiting-time constrained event graphs with links of weights $\Delta t \leq \delta t$ removed contain all $\delta t$-limited paths. (d) Reduced event graph in which locally redundant links are removed (see main text). The highlighted line represents a time respecting path (a) and its equivalent path over event graph (b,c) and reduced event graph (d).}
    \label{fig:schematic-event-graph}
\end{figure}

\paragraph{Modelling approach.}
A temporal network $G = (\mathcal{V}, \mathcal{E},\mathcal{T})$ is defined as a set of nodes $\mathcal{V}$ connected through events $e=(u,v, t_\text{start}, t_\text{end}) \in \mathcal{E}$, each of which represents an interaction of two nodes $u, v \in \mathcal{V}$ starting at time $t_\text{start}$ and ending at time $t_\text{end}$ observed during an observation period $\mathcal{T}$ (i.e., $t_\text{start}, t_\text{end} \in \mathcal{T} \,\forall_{e} \in \mathcal{E}$ and $t_\text{start} < t_\text{end}$). The connectivity of events is characterized by time-respecting paths \cite{holme2005network,lentz2013unfolding}, defined as sequences of adjacent events. Here we call two distinct events $e, e' \in \mathcal{E}$ adjacent, and denote this by $e \rightarrow e'$, if they follow each other in time ($t'_\text{start} > t_\text{end}$) and share at least one node in common ($\{ v, u \} \cap \{v', u'\} \neq \emptyset$) as demonstrated in Fig. \ref{fig:schematic-event-graph}a. For simplicity, we assume that temporal network events are instantaneous ($t_\text{start} = t_\text{end}$), but all of our notations can be easily extended to directed events and to temporal hypergraphs \cite{badiemodiri2020efficient,mellor2019event}.

While time-respecting paths encode the possible routes of information, some dynamical processes have further restrictions on the duration they can propagate further after reaching a node. For example, in disease spreading, infected nodes may recover after some time, becoming unable to infect other nodes unless re-infected. In our definition, we define limited-waiting-times in temporal paths by allowing adjacent events $e=(u, v, t_\text{start}, t_\text{end})$ and $e'=(u', v', t'_\text{start}, t'_\text{end})$ to be connected ($\delta t$-adjacent) only if there is less than $\delta t$ time between them (i.e., $t'_\text{start} - t_\text{end} < \delta t$). In contrast to the control parameters based on node or event occupation probabilities, which could be used to adjust the overall activity level of the network, changing $\delta t$ modifies the behavior of the spreading itself. Additionally, processes unconstrained by waiting time can be modeled as a special case of the limited waiting-time process, with an infinitely large value of $\delta t$.

A compact way of describing the problem of reachability on temporal networks is provided by weighted event graph representation $D = (\mathcal{E}, E_D, \Delta t(e, e'))$, a static directed acyclic representation of temporal networks~\cite{kivela2018mapping}. In this description events act as nodes and two events $e$ and $e'$ are connected through a directed, weighted link if they are adjacent with weights defined as $\Delta t(e, e') = t'_\text{start} - t_\text{end}$, i.e.,
$E_D = \{ (e, e') \in \mathcal{E} \times \mathcal{E} \ |\  e \rightarrow e' \}$. The event graph contains a superposition of all temporal paths \cite{saramaki2019weighted} and retains the arrow of time even after turning the temporal structure into a static one (Fig.~\ref{fig:schematic-event-graph}b). Event graph representation of temporal networks has proven to be suitable for studying properties of temporal networks such as occurrences of motifs~\cite{kovanen2011temporal}, decomposition of the temporal network into smaller components~\cite{mellor2018analysing} and providing a lower-dimensional embedding of the temporal network that can be consumed by many machine learning method~\cite{torricelli2020weg2vec}. For our use case, a superposition of all $\delta t$ limited-time temporal paths ($D_{\delta t}$) of the temporal network can be achieved by constructing the event graph of the temporal network and removing all the event graph links with weights larger than $\delta t$, in other words, $D_{\delta t}$ is a directed graph with the same set of vertices and the same weight function as $D$ and set of edges $\{(e, e') \in E_D\ |\  \Delta t(e, e') \leq \delta t\}$ (see Fig.~\ref{fig:schematic-event-graph}c).

Furthermore, we define the reduced temporal event graph $\hat{D}$ and its waiting-time constrained variation $\hat{D}_{\delta t}$, where only the first adjacency relationships per temporal network node for each event are retained. $\hat{D}$ and $\hat{D}_{\delta t}$ nodes have a maximum in- and out-degree of two, yet they contain all the reachability relationships of the original event graph \cite{Mellor}. That is, the reduced event graph exactly retains the reachability of the original event graph by removing redundant connections (feed-forward loops) between events. The reduction allows interpretation of the three possible out-degrees using the terminology of directed percolation as annihilation (0), diffusion (1), and decoagulation (2) in the case the out-neighbors are not already reachable through some longer loop. Note that this upper bound on in- and out-degrees is valid if the probability of simultaneous occurrence of adjacent events is negligible. See Supplementary Materials (SM) for more details.

\paragraph{Order parameters and other characteristics.}
Compared to static structures, temporal networks incorporate time as an additional degree of freedom, which introduces an extra dimension to the characterization of their structural phase-transition of connectivity around a critical point. This is similar to directed percolation where dimensions are related to space and time with associated independent critical exponents \cite{1982ZPhyB..47..365G, 1981ZPhyB..42..151J}. We measure the expected $\delta t$-limited waiting-time reachability starting from a random event $e$. Of interest is the number of unique reachable nodes $ \mathcal{V}_{e \rightarrow}\subseteq \mathcal{V}$, the time duration of the longest path (i.e., its lifetime \cite{kivela2018mapping}) $\mathcal{T}_{e \rightarrow}\subseteq \mathcal{T}$, and the total number of reachable events $\mathcal{M}_{e\rightarrow}\subseteq \mathcal{E}$. The expected values of these are analogous to mean spatial volume $V= \langle | \mathcal{V}_{e \rightarrow}| \rangle $, mean survival time $T= \langle \max \mathcal{T}_{e \rightarrow} - \min  \mathcal{T}_{e \rightarrow} \rangle $, and mean cluster mass $M= \langle | \mathcal{M}_{e \rightarrow}| \rangle $ in the directed percolation formalism (respectively) \cite{henkel2008non,hinrichsen2000non}. Further, in parallel to directed percolation, we define the survival probability $\surv{t}$ as the probability that there is a path from a randomly selected initial source event at $t_0$ to an event after time $t_0 + t$. The ultimate survival probability $\usurv=\lim_{t \to \infty} \surv{t}$ is then the survival probability at large values of $t$. Note that when defining these quantities we opted for simplicity (see SM for discussion).

Using the maximum waiting-time $\delta t$ as a control parameter is a natural choice as it has a clear physical interpretation. However, unlike occupation probabilities that are typically used as control parameters in directed percolation, the scale of $\delta t$ depends on the timescales of the system. Further, although it is related to the local connectivity, this relationship is indirect and might depend on, e.g., the temporal inhomogeneities in interaction sequences. For this reason, we define another control parameter that directly measures the local connectivity of the system. We use the local effective connectivity $\expqdegout(\delta t)$, which is the average excess out-degree of the reduced event graph $\hat{D}_{\delta t}$. This is a monotonically increasing function of $\delta t$, which normalizes the changes in connectivity given by the changes in the maximum allowed waiting-time $\delta t$. We then centralise this quantity by subtracting its value from its phase-transition critical point $\expqdegout_c$, and denote the resulting control parameter as $\tau = \expqdegout - \expqdegout_c$. 

In addition to the single-source scenario, where the component starts from a single node in $D_{\delta t}$, we investigated the fully-occupied homogeneous initial condition, where we compute paths starting from all nodes in $D_{\delta t}$ with time $t < t_0$. Analogous to directed percolation, we define particle density $\rho(t)$ as the fraction of infected nodes in $D_{\delta t}$ at time $t$, while stationary density $\rho_\text{stat}(\tau)$, the order parameter, is defined as the particle density after the system reached a stationary state. We can incorporate the effects of an external field $h$ to this scenario: in continuous-time, this would be equivalent to the spontaneous emergence of sources of infection, i.e.~occupation, of nodes in $D_{\delta t}$ (events in $G$) through an independent Poisson point process with rate $h$. Susceptibility $\chi(\tau, h)=\frac{\partial}{\partial h} \rho_\text{stat}(\tau, h)$ can then be measured through observing the effect of changing the external field \cite{henkel2008non}.

\begin{figure}[t!]
    \centering
    \includegraphics[width=1\linewidth]{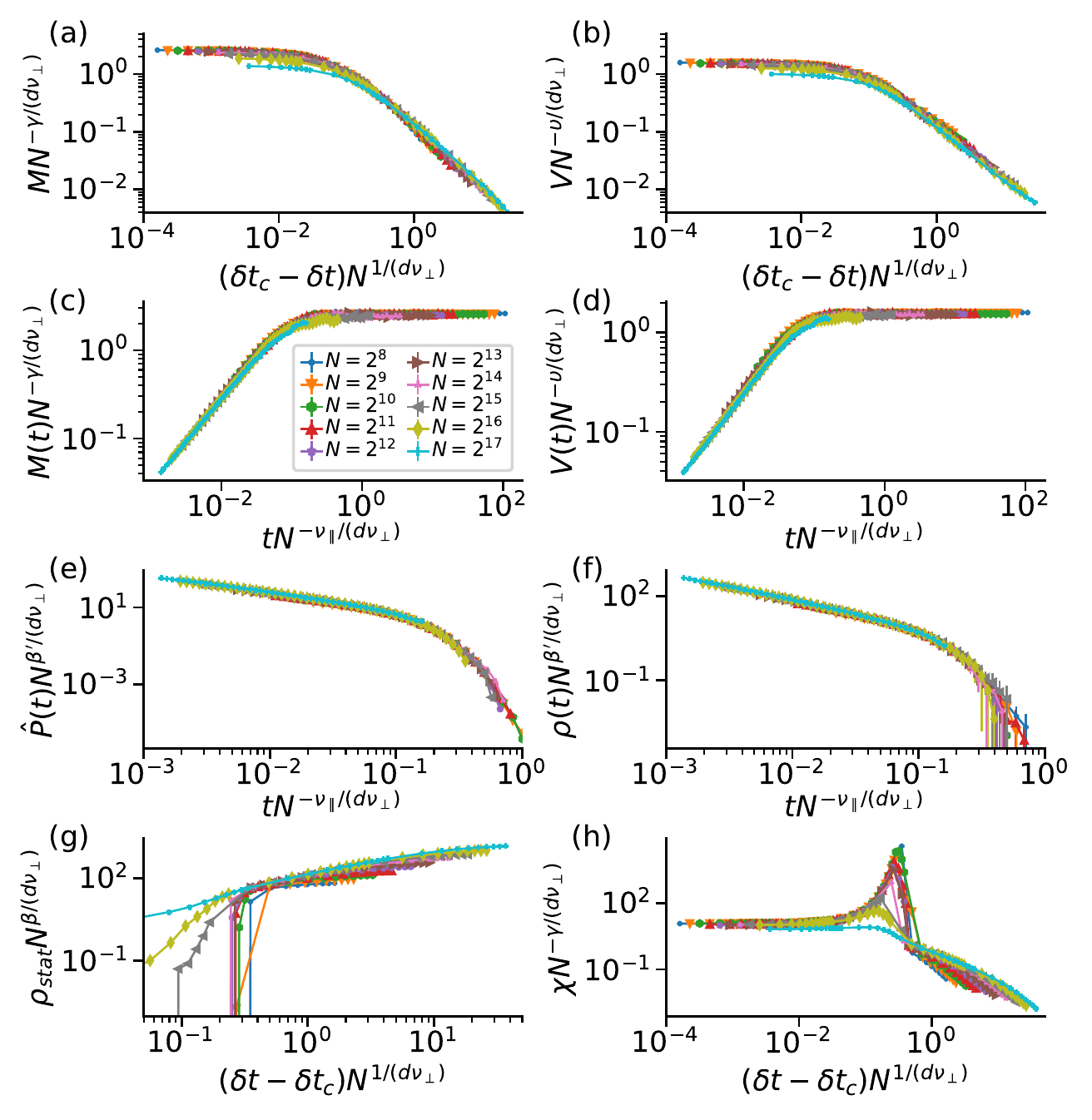}
    \caption{finite-size scaled (a,c) Mean cluster mass $M$, (b,d) volume $V$ and (e) survival probability $\hat{P}(t)$ for single-source spreading scenarios. (f) Particle density $\rho(t)$, (g) static density $\rho_\text{stat}$ and (h) susceptibility $\chi(\delta t, 0)$ as a function of $\delta t$ for the homogeneous initial condition. Measurements are averaged over at least 256 (up to 4096) realisations of temporal network constructed from random 9-regular networks ($N\in\{2^8,\ldots,2^{17}\}$) and Poisson point process activations $\lambda=1$ of links. All functions of time are measured at $\delta t=\delta t_c=0.08808$. $d$ is set to directed percolation upper critical dimension $d_c = 4$.
    }
    \label{fig:finite-size-scaling}
\end{figure}

\paragraph{Critical behavior in random systems.}
Next, we derive a mean-field approximation for the above-defined measures and identify the critical point. We model temporal networks with an underlying static structure, where events are induced via links activating by independent and identical continuous-time stochastic processes. In order to do so, we need to first derive the degree distribution of the reduced event graph $\hat{D}_{\delta t}$, i.e. probabilities that one can reach zero, one, or two events from a randomly chosen event in the temporal network.
Given the excess degrees $\ulqdegval{l}$ and $\ulqdegval{r}$ of the two temporal network nodes in $G$ incident to the link corresponding to the event $e\in \mathcal{E}$, we can compute the probability of a zero out-degree for a node in $\hat{D}_{\delta t}$ (i.e., an event in original temporal network $G$) as $\pdegout{0}=\Pi_{\delta t}\hat{\Pi}_{\delta t}^{\ulqdegval{l}+\ulqdegval{r}}$. Here $\Pi_{\delta t}$ is the cumulative inter-event time distribution induced by a link activation process for a given $\delta t$, and $\hat{\Pi}_{\delta t}$ is the corresponding cumulative residual inter-event time distribution. Similarly, for out-degree $2$, we can compute
\begin{equation}
\pdegout{2}=\int_0^\infty (1- \hat{\Pi}_{\min{\delta t, t}}^{\ulqdegval{l}})(1- \hat{\Pi}_{\min{\delta t, t}}^{\ulqdegval{r}})\pi_t dt,    
\end{equation}
where $\pi_t$ is the inter-event time distribution. Given that the maximum out-degree of events in the reduced event graph is $2$, the $\pdegout{1}$ can be derived as $\pdegout{1}=1-\pdegout{0}-\pdegout{2}$. In-degree probabilities can be derived similarly.

The joint in- and out-degree distribution of the event graph can be computed from the excess degree distribution $\ulqdeg{k}$ of the underlying static network. If the degrees are independent, this becomes $\pdeginout{i}{o} = \sum_{\ulqdegval{l},\ulqdegval{r}} \pdegin{i} \pdegout{o} \ulqdeg{l} \ulqdeg{r}$. We will denote the generating function of the joint degree distribution as $\mathcal{G}_0(z_{\text{in}},z_{\text{out}})$ and the corresponding out excess degree distribution as $\mathcal{G}_1^{\text{out}}(z_{\text{out}})$. We construct the mean-field rate equation for occupation density $\rho(t)$ in homogeneous occupation initial condition using the excess out-degree distribution of the event graph $\qdegout{k} = \frac{d^k}{k! d z^k} G^{out}_1(z) |_{z=0}$. The excess out-degree of nodes in the event graph $\hat{D}$ gives the change in the number of further nodes we can reach from an already reached node: nodes with out-degree 2 increase the number of reached nodes by one, nodes with out-degree 1 do not affect on the number of reached nodes, and out-degree 0 nodes reduces by one the number of reached nodes. The total change, therefore, is $\qdegout{2} - \qdegout{0}$. In addition, some nodes we can reach are already reachable through other paths. In total we reach on expectation $\qdegout{1}  + 2\qdegout{2}$ nodes where each node is already reached with probability $\rho(t)$. The rate equation becomes
\begin{equation}\label{eq:rate-equation}
    \partial_t \rho(t)= [\qdegout{2} - \qdegout{0}]\rho(t) - [\qdegout{1}  + 2\qdegout{2}]\rho^2(t)\,.
\end{equation}
In this equation the values of $\qdegout{k}$ are constants in time. Noting the critical point for this equation as $\qdegout{2} - \qdegout{0}=0$, and that the expected value is by definition $\expqdegout =\qdegout{1}  + 2\qdegout{2}$, and that $\qdegout{2} - \qdegout{0} = \expqdegout - 1$, we can write Eq.~\eqref{eq:rate-equation} as $\partial_t \rho(t) = \tau \rho(t) - \expqdegout \rho^2(t)$.

Equation \eqref{eq:rate-equation} follows the same form as the directed percolation mean-field equation for a $d+1$-dimensional lattice \cite{henkel2008non} and can be solved explicitly (see SM). It has the critical point at $\tau = 0$, while it indicates that $\rho \rightarrow \tau/\expqdegout $ for $\tau > 0$. Asymptotically it provides the critical exponents as $\rho(t) \sim t^{-\alpha}$ at $\tau=0$ and $\rho_\text{stat}(\tau) \sim \tau^{\beta}$ when $\tau > 0$ and $t \rightarrow \infty$ with values $\alpha=\beta=1$, where $\alpha = \beta/\nu_\parallel$ and $\nu_\parallel$ is the temporal correlation length exponent, in accordance with the corresponding mean-field directed percolation critical exponents~\cite{henkel2008non}.

The expected out-component size, i.e.~mean cluster mass $M$, can be computed from the joint degree distribution of the event graph $\hat{D}_{\delta t}$ by assuming that it is a random directed graph with the same joint in- and out-degree distribution as $\hat{D}_{\delta t}$. The out-component size distribution probability-generating function $H_0$ can be derived from
\begin{equation}
\begin{aligned}
H_0(z_{\text{out}}) &= z_{\text{out}}\mathcal{G}_0(1,H_1(z_{\text{out}}))\\
H_1(z_{\text{out}}) &= z_{\text{out}}\mathcal{G}^{out}_1(H_1(z_{\text{out}})),    
\end{aligned}
\end{equation}
and the mean out-component size can be written as $M = \frac{\partial H_0(z_{\text{out}})}{\partial z_{\text{out}}} |_{z_{\text{out}}=1}$ \cite{newman2001random}. These equations, when $\tau \rightarrow 0^-$, lead to $M \sim -\tau^{-\gamma}$ with $\gamma=1$ (see SM). Here $\gamma=\nu_\parallel + d \nu_\perp - \beta - \beta'$, matching the mean-field exponent of mean cluster mass in directed percolation \cite{henkel2008non}. Here, $\nu_\perp$ indicates the spatial temporal correlation exponent.

The component survival probability, $\surv{t}$, is measured by the out-component time-span of nodes in the event graph and the occupation density, $\rho(t)$, is calculated by the in-component sizes of all possibly reachable nodes, implying that these two quantities are equal ${\rho(t)=\surv{t}}$ (see SM). Consequently, given the control parameter $\tau$, $\rho_\text{stat}(\tau) = \usurv(\tau)$ as long as the time-reversed event graph has the same probability of being generated as the original one (e.g., if $\forall_{i,o} \, p^{in,out}_{i,o} = p^{in,out}_{o,i}$). This leads us to the rapidity-reversal symmetry for event graphs similarly characterizing directed percolation \cite{grassberger1979reggeon} where $\beta = \beta'$ and $\usurv(\tau) \sim \tau^{\beta'}$. Note that while the condition above holds for a variety of random temporal network models, for real-world systems intuition might suggest e.g.~a higher probability of $p^{in,out}_{1,2}$ as compared to $p^{in,out}_{2,1}$ due to over-representation of causal motifs \cite{kovanen2011temporal}. In practice, however, we observed no deviations from the above condition in two large real-world system (see SM).

\paragraph{Finite-size scaling in random systems.}
The critical exponents can be empirically verified through finite-size scaling of the system close to its percolation critical point, where its large-scale properties become invariant under scale transformations. We simulate random temporal networks of varying size and perform efficient reachability estimations \cite{badiemodiri2020efficient} from single-source and homogeneous fully-occupied initial conditions. We expect that curves of macroscopic quantities collapse when using the correct critical exponents of $\beta$, $\nu_\parallel$ and $\nu_\perp$ corresponding to the mean-field values of directed percolation. The results confirm that the directed percolation mean-field exponents characterize the percolation phase transition of random temporal networks. This is demonstrated in Fig.~\ref{fig:finite-size-scaling}a-f for temporal networks induced on a 9-regular network with links activated via independent Poisson processes. These results are robust in the presence of several types of temporal and spatial heterogeneities~\cite{longpaper}.

\paragraph{Directed percolation measures in real-world temporal networks.}
\begin{figure}[ht!]
    \centering
    \includegraphics[width=1\linewidth]{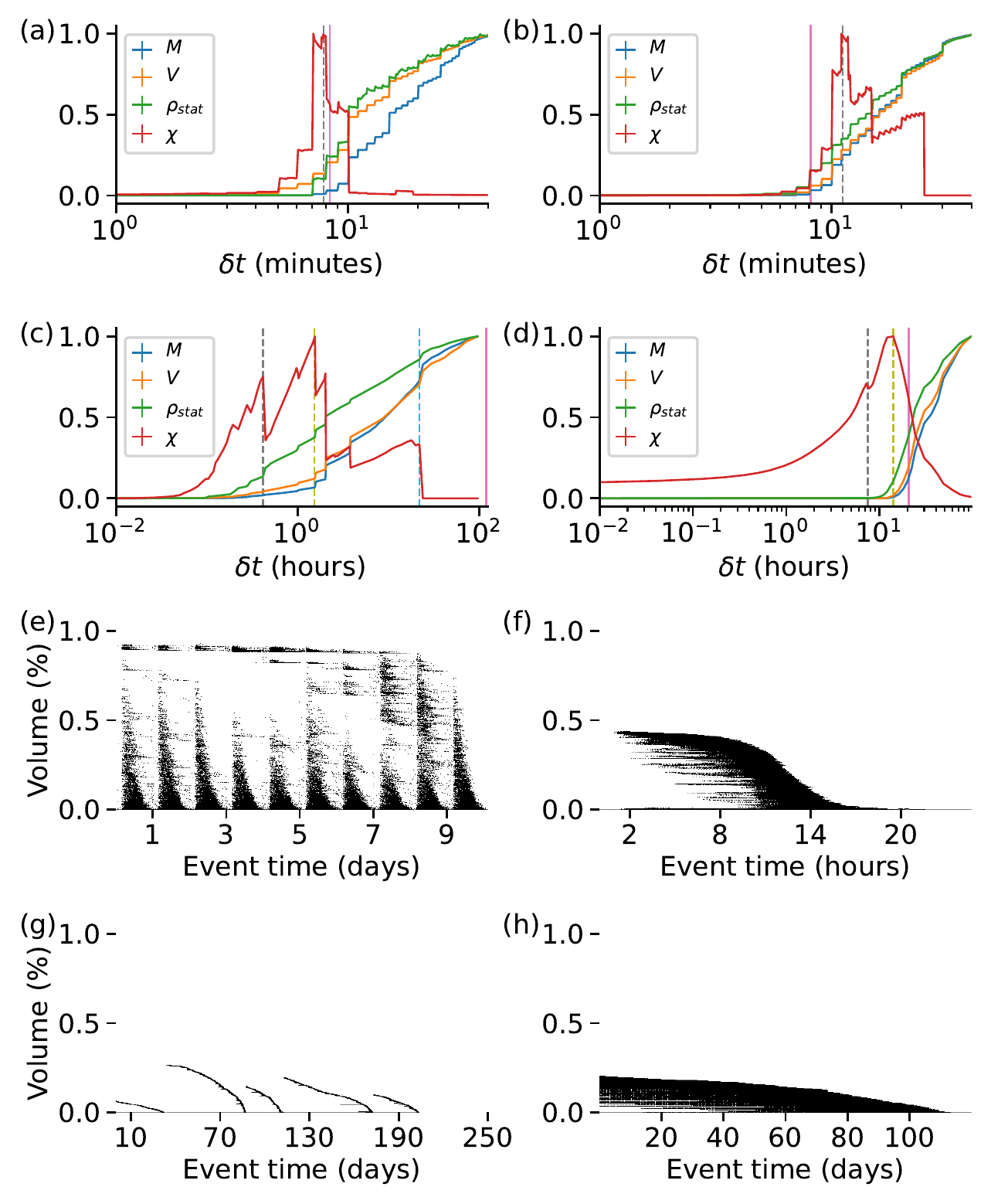}
    \includegraphics[width=1\linewidth]{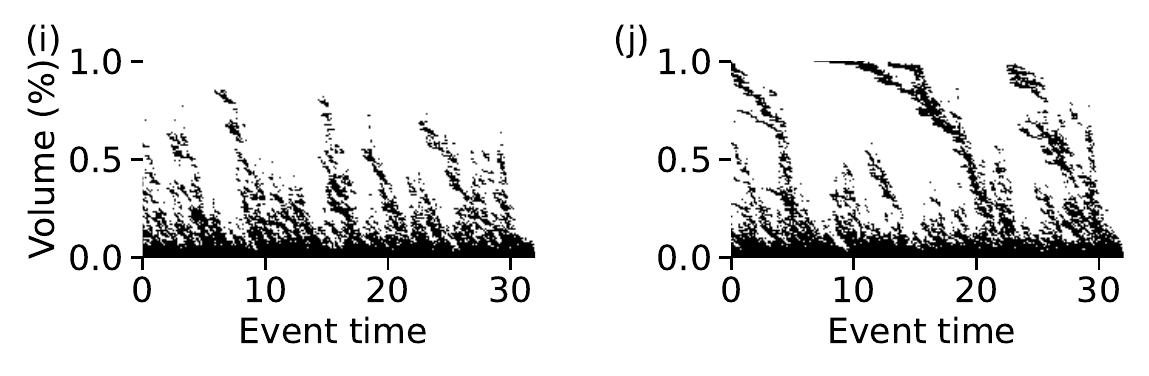}
    \caption{Mean cluster mass $M$, mean cluster volume $V$, static density $\rho_\text{stat}$ and susceptibility $\chi(\delta t, 0)$ as a function of $\delta t$ for four real-world networks: (a) Air transport \cite{bts2017air}, (b) Helsinki public transportation \cite{kujala2018collection}, (c) Twitter mentions \cite{yang2011patterns} and (d) mobile phone calls \cite{karsai2011small} display an absorbing to active phase transition around 470 seconds, 670 seconds, 25 minutes and 7.5 hours respectively, as indicated by change from very small values for $M$, $V$ and $\rho_\text{stat}$ to values comparable to the size of the system and a peak in susceptibility $\chi(\delta t, 0)$. Mobile and Twitter networks show a second peak in susceptibility around 1.5 hours and 22 hours, respectively, and Twitter data shows a third peak around 14 hours. The trajectories are re-scaled to the range $[0,1]$.
    $\delta t_c$ is estimated using the analytical solution from Ref.~\cite{longpaper} by approximating the network to temporal network with a random regular static base and Poisson point process activation. This estimates the threshold respectively at 500 seconds, 488 seconds, 119.1 hours and 22.5 hours, displayed using solid vertical lines on a-d. The temporal reachability profiles display relative cluster volumes for each event as a function of the event time for $\delta t \approx \delta t_c$ for (e) air transport, (f) Helsinki public transportation, (g) Twitter mentions and (h) mobile phone call networks. The reachability profiles for random 9-regular network of size 1024 with Poisson point process $\lambda=1$ and (i) $\delta t  = 0.088 \approx \delta t_c$ and (j) $\delta t = 0.092$.}
    \label{fig:real-world-phase-transition-sixteenth}
\end{figure}

We measure the same macroscopic quantities as before for four different real-world systems, concentrating on temporal networks describing air transportation, public transportation, Twitter mentions, and mobile phone calls (Fig.~\ref{fig:real-world-phase-transition-sixteenth}a-d respectively). In these networks, an event represents respectively: a flight between two airports in the United States, a public transport vehicle transiting between two consecutive stations on a typical Monday in Helsinki, a user mentioning another user in a tweet on Twitter and a mobile phone subscriber calling another subscriber of a major European carrier. For details on the data sets, see Tab.~\ref{tab:real-world-sizes} and SM. In each system, there is clear evidence of an absorbing to active phase transition in terms of $M$, $V$, and $\rho_\text{stat}$. Note that the scales of these quantities are not directly comparable, highlighting the fact that distinguishing between the different notions of connectivity is important in practical terms. Further, multiple peaks in susceptibility indicate multiple connectivity time scales.

\begin{table}[h!]
\caption{\label{tab:real-world-sizes} The composition of the real-world temporal networks studied. Note that the Air and Public transport networks are directed temporal networks with non-instantaneous events, meaning that each event (a flight or a trip between consecutive stations) has a different start time and end time ($t_\text{start} < t_\text{end}$) corresponding to departure and arrival time of the vehicle. On the other hand, the Twitter and Mobile datasets were constructed as undirected, instantaneous temporal networks to model bi-directional information flow between the users/subscribers.}
\begin{ruledtabular}
\begin{tabular}{cll}
\multicolumn{1}{c}{\textrm{Dataset}} &
\multicolumn{1}{c}{Nodes} &
\multicolumn{1}{c}{Events}\\
\colrule
Air Transport~\cite{bts2017air} & 279 airports & 180\,112 flights \\
Public Transport~\cite{kujala2018collection} & 6858 stations & 664\,138 trips \\
Twitter~\cite{yang2011patterns} & 17\,313\,552 users & 266\,179\,671 mentions \\
Mobile~\cite{karsai2011small} & 5\,193\,086 users & 324\,576\,400 calls \\
\end{tabular}
\end{ruledtabular}
\end{table}

The reachability phase transition can be better understood by investigating temporal connectivity profiles represented by cluster volumes of individual events. Structures similar to those of random networks (see SM) can be observed for Air Transport and Twitter (Fig.~\ref{fig:real-world-phase-transition-sixteenth}e,g). However, in Air Transport, the structure is regular, following the diurnal pattern of flights. In Twitter, the components do not reach most nodes due to the greater separation of temporal components, and their structure reflects the rare emergence of possible macroscopic cascades. Public Transport (one day) and Mobile networks display a single wing-like structure (Fig.~\ref{fig:real-world-phase-transition-sixteenth}f,h). This is induced by early components that can reach a significant fraction of nodes, which are then joined by other components reaching smaller subsets. This is also indicated by the horizontal structures under the wings.

\paragraph{Conclusion.}
The connectivity of a network is an important measure of its resilience and an underlying concept for any dynamical process running on it. It encodes the possible transportation routes or paths of information diffusion and determines how misinformation or diseases spread in real-world settings. The connectivity of static networks and related dynamical processes are routinely analyzed within the framework of (isotropic) percolation theory \cite{newman2002spread, barrat2008dynamical, pastor2015epidemic} with methods borrowed from critical phenomena \cite{hinrichsen2000non, dorogovtsev2008critical}. Furthermore, many natural or synthetic networks, ranging from the brain \cite{chialvo2010emergent,hesse2014self} or artificial neural networks \cite{shin2006self} to geological phenomena \cite{bak1989earthquakes} and urban systems \cite{chen2008scaling} tend to self-organize their medium or their parameters or be optimized by outside intervention towards criticality \cite{bak1987self,bak1988self}. Therefore, it is of great utility to locate the onset of critical phase-transitions points and predict the behavior of the system in that vicinity.

While connectivity transitions and the critical behaviour of the system are understood in static networks by means of isotropic percolation theory, temporal networks, by and large, have been out of reach of a similar methodology. This has practical implications as connectivity is a limiting factor of any dynamical processes and at the same time temporal interactions have been shown to have dramatic effects on the speed and volume of any ongoing dynamical process~\cite{lambiotte2016guide,holme2012temporal,holme2015modern}. For example, disease spreading in static networks can be mapped to a percolation process leading to a theoretical understanding of the epidemic threshold as a consequence of connectivity phase transition \cite{newman2002spread}. This connection has been extensively exploited to use the mathematical machinery of network percolation to derive various theoretical and practical results on static networks \cite{pastor2015epidemic, blasius2020power}. In temporal networks, such analysis is typically based on theoretical results on sequences of static networks \cite{leitch2019toward} or case studies based purely on simulations \cite{karsai2011small,barrat2021effect}. The concise theory of temporal network connectivity provided here shows that the reachability phase transition in temporal networks belongs to the directed percolation universality class, which is a necessary step forward from the limited description provided by the theory of static networks. It also indicates that directed percolation may have many counterparts in reality with the expected scaling relations.

The mapping presented in this paper allows for predicting the critical thresholds and the connectivity behaviors of a diverse set of systems that can be modeled as temporal networks. Now, similar to static network connectivity, not only we have theoretically grounded summary statistics of the component size distribution (the order parameters and cluster mass, volume and lifetime), but also we know ways to find their transitions even in finite-size systems. Moreover, we now possess a theory to predict the behavior of such random systems and find transition points accurately. Real networks are often approximated with random graphs, and the random models are used as reference points: deviations from the minimal random models expose important structural features of the real systems, and conversely, agreement with these models tells that the structures, correlations, and inhomogeneities present in the data do not have a measurable effect on the connectivity. Although introduction of heterogeneities might shift the critical threshold of connectivity in temporal networks, the directed percolation phase transition is surprisingly robust to several types of temporal and topological heterogeneities~\cite{longpaper}. Consequently further research is required to find the boundaries and extremities of application of this framework on theoretical and real-world networks.

\begin{acknowledgments}
\paragraph{Acknowledgements.}
We would like to thank János Kertész and Géza Ódor for their helpful comments and suggestions. The authors wish to acknowledge CSC -- IT Center for Science, Finland, and Aalto University ``Science-IT'' project for generous computational resources. Márton Karsai acknowledges support from the DataRedux ANR project (ANR-19-CE46-0008) and the SoBigData++ H2020 project (H2020-871042).
\end{acknowledgments}

\bibliography{citations}
\end{document}


\title{Supplementary Material for Directed Percolation in Temporal Networks}
\author{Arash Badie-Modiri}
\affiliation{Department of Computer Science, School of Science, Aalto University, FI-0007, Finland}
\author{Abbas K.~Rizi}
\affiliation{Department of Computer Science, School of Science, Aalto University, FI-0007, Finland} 
\author{Márton Karsai}
\affiliation{Department of Network and Data Science
Central European University, 1100 Vienna, Austria}
\affiliation{Alfr\'ed R\'enyi Institute of Mathematics, 1053 Budapest, Hungary}
\author{Mikko Kivelä}
\affiliation{Department of Computer Science, School of Science, Aalto University, FI-0007, Finland} 

\date{\today}

\maketitle
\section{Implementation}
The implementation, along with two of the real-world temporal networks used, namely US air transport and Helsinki public transport, are made available online \cite{badie2021implementation}. Please refer to Supplementary Material for \cite{longpaper} for a more detailed usage information.

\section{Generating synthetic temporal networks and the event graph}
The synthetic temporal networks are created with some continuous-time stochastic process based on an underlying static network with a degree distribution of $p_k$ and excess degree distribution of $q_k$. The event graph, directed acyclic graph of adjacency relationships between pairs events, can then be produced by iterating through all events $e$ and connecting it to all other events when $e$ happens less that $\delta t$ time before that event and they share at least one node.

Reachability on the event graph will be preserved by removing some of the links so that the in/out-degree varies between 0 to 2 for every node \cite{Mellor} as long as the probability of adjacent events happening at exactly the same time is negligible. Practically, for every event $e$ in the event graph, we can remove directed links to all but the very first events for each of the two nodes involved in $e$. This preserves connectivity in the event graph since all the events on the other end of the removed adjacency relationships would still be connected through one of the remaining links out-bound from $e$ as they share at least one node and the time difference is less than or equal to the original event. Note that if more than one adjacent events are happening at the same time and no other adjacent events happen before them, we would have to keep all of them to preserve connectivity.

Note that in practice it is often not necessary to explicitly generate the event graph to measure the quantities. It is possible to store the list of associated events for each node in the network sorted by time and generate adjacency relationships on the fly. This can also be combined with other techniques such as using probabilistic data structures for estimating out-component sizes to allow processing of temporal networks much larger than what is possible with the explicit solution \cite{badiemodiri2020efficient}. 

\subsection{Degree distribution of the reduced event graph}
\begin{figure}
    \centering
    \includegraphics[width=1.0\linewidth]{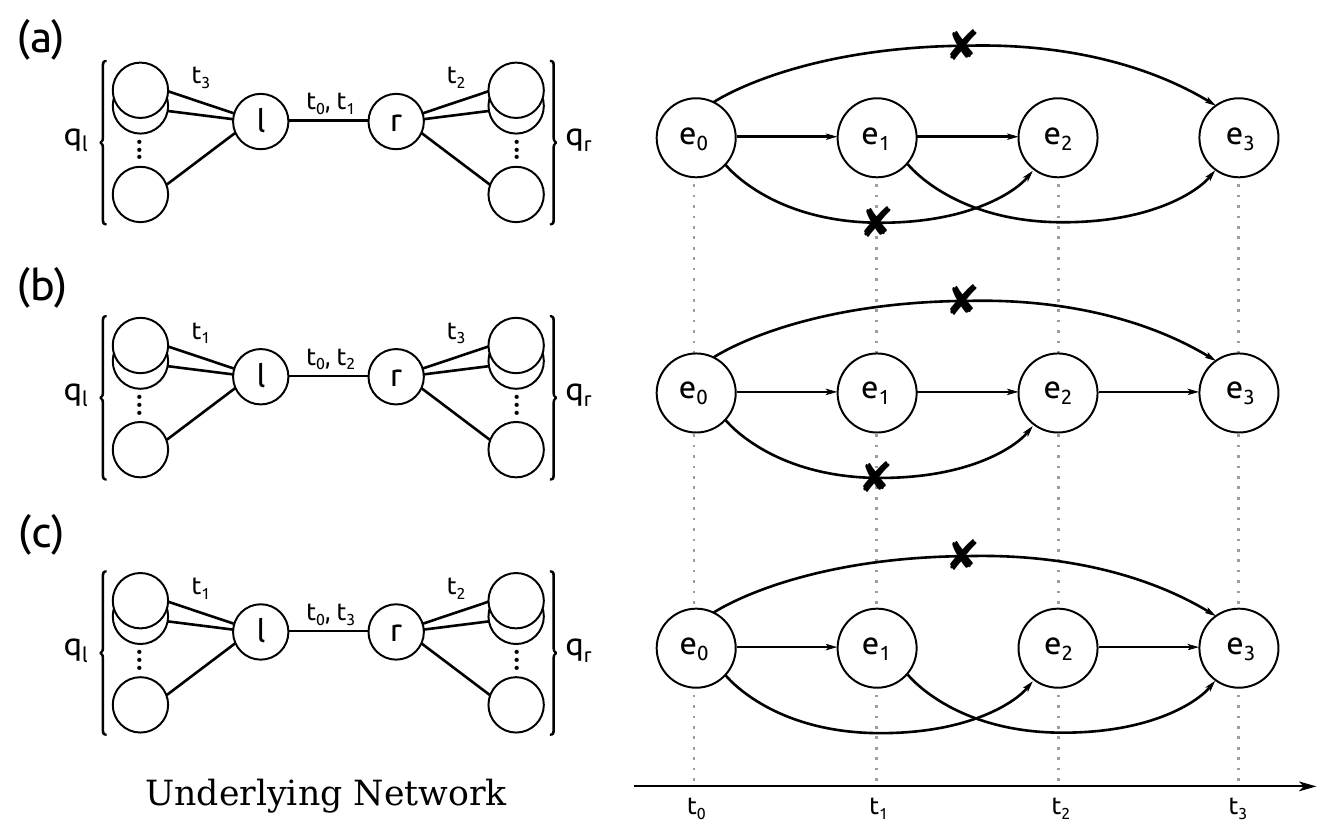}
    \caption{Considering the case of an event between nodes $l$ and $r$ happening at time $t_0$, where each node has $q_l$ and $q_r$ neighbours other than each other respectively. Assuming link ${l \undirectededge r}$ was selected uniformly at random from the set of all the links in the base network, the values $q_l$ and $q_r$ are both realisations of the excess degree distribution of the base network $P_q$. Out-degree of the event $e_0 = (l, r, t_0)$ is between zero and two depending on the order and timing of events between $l$, $r$ and their neighbours. If the ${l \undirectededge r}$ link activates before any of the other links incident to $l$ and $r$ (panel a) or only links incident to $l$ (or only $r$) other than ${l \undirectededge r}$ fire before ${l \undirectededge r}$ (panel b) at a time $t_1 > t_0$, event $e_0$ would have an out degree of zero if $t_1 - t_0 \geq \delta t$ or one if $t_1 - t_0 < \delta t$. All other edges coming out of $e_0$ would necessarily get pruned out as shown by the crossed-out links. The only case for $e_0$ having a degree two happens when at least one event at $t_1 < \delta t$ only involving $l$ and not $r$ and one at $t_2 < \delta t$ only involving $r$ and not $l$ both happen before ${l \undirectededge r}$ fires again.}
    \label{fig:schematic_with_mellor}
\end{figure}

Let's assume a vertex on the event graph, an event $e$, that involves two nodes called $l$ and $r$ which just activated at time $t_0$ (Fig \ref{fig:schematic_with_mellor}). The two nodes $l$ and $r$ have respectively $q_l$ and $q_r$ neighbor nodes, other than each other, over the static network.

Let's also define $Pr(t_{res} < \delta t)$ as the probability that a process with inter-event time distribution $\mathcal{T}$ can activate at least once in time $\delta t$ after a random point in time. This can correspond to probability of one of the links in the underlying network activating within a time period of $\delta t$. Random variable $t_{res}$ is distributed according to the residual inter-event time distribution $\mathcal{R}$. Similarly, $Pr(t_{iet} < \delta t)$ is the probability that a process with inter-event time distribution $\mathcal{T}$ can activate at least once in time $\delta t$ right after activation.

Probability of an event having out-degree of zero in the event graph can be calculated as:
\begin{equation}
    P_{out}(0 | q_l, q_r) = Pr(t_{res} > \delta t)^{q_l + q_r} Pr(t_{iet} > \delta t)
\end{equation}
where $q_l$ and $q_r$ are the number of neighbours each of the nodes participating in the event has except for the connection between two nodes of the event in question, $t_{iet}$ is a realisation of the inter-event time distribution of the network $\mathcal{T}$ and $t_{res}$ is a realisation of the residual inter-event time distribution $\mathcal{R}$. Out-degree of an event is zero if and only if none of the $q_l + q_r$ adjacent links on the underlying network have an event within $\delta t$ and the two nodes participating in the original event also don't have any events between them within $\delta t$. The second term corresponds to the probability of the same link not activating and the first is the probability of all of the other incident links except for the original link not activating in $\delta t$.

The only case that an event on the event graph can have an out-degree equal to 2 (as shown on Fig. $\ref{fig:schematic_with_mellor}c$) is that at least one of the $q_l$ neighbours of $l$ and one of the $q_r$ neighbours of $r$ activate before $\delta t$ and before reactivation of the link between $l$ and $r$. Activation of the link between $l$ and $r$ before at least one of the links on each side is activated (Fig. $\ref{fig:schematic_with_mellor}a$ and $\ref{fig:schematic_with_mellor}b$) would result in out-degree equal to zero or one depending on the value of $\delta t$ and timing of the events.

Probability of having an out-degree equal to 2 can be calculated this way:
\begin{equation}
\begin{aligned}
    P_{out}(2 | q_l, q_r) = \int^\infty_0 &(1 - Pr(t_{res} > \delta t \lor t_{res} > t)^{q_l} )\\
                                           &(1 - Pr(t_{res} > \delta t \lor t_{res} > t)^{q_r})\\
                                           &Pr(t \sim \mathcal{T}) \dd{t}
\end{aligned}
\end{equation}
where $t_{res}$, $\mathcal{T}$, $q_l$ and $q_r$ are defined as above. An event has an out-degree equal to 2 if and only if two mutually non-adjacent links adjacent to the link corresponding to the original event activated within $\delta t$ and before the link corresponding to the original event is activated.

\begin{equation}
    P_{out}(1 | q_l, q_r) = 1 - (P_{out}(0 | q_l, q_r) + P_{out}(2 | q_l, q_r))
\end{equation}

Based on these equations, it is trivial to construct joint in- and out-degree distribution
\begin{equation}\label{eq:joint-degree-dist}
\begin{aligned}
    P(in, out) = \sum^\infty_{q_l,q_r=1} P_{in}(in | q_l, q_r) P_{out}(out | q_l, q_r) \\
    P_q(q_l) P_q(q_r)
\end{aligned}
\end{equation}
where $P_q(i)$ is the probability mass function of excess degree for the static aggregate base network.

It is possible to construct the joint degree distribution generating function $\mathcal{G}$ using the joint degree distribution itself
\begin{equation}\label{eq:joint-degree-dist-generating-fuction}
    \mathcal{G}(x,y) = \sum^2_{in, out = 0}P(in, out) x^{in} y^{out}
\end{equation}
and in- and out-degree and excess degree distribution generating functions
\begin{equation}\label{eq:generating-fuction-magic}
\begin{aligned}
    \mathcal{G}^{in}_0(x) &= \mathcal{G}(x, 1)\\
    \mathcal{G}^{out}_0(y) &= \mathcal{G}(1, y)\\
    \mathcal{G}^{in}_1(x) &= \frac{1}{z} \frac{\partial \mathcal{G}(x,y)}{\partial y}\biggl\vert _{y=1}\\
    \mathcal{G}^{out}_1(y) &= \frac{1}{z} \frac{\partial \mathcal{G}(x,y)}{\partial x}\biggl\vert _{x=1}\\
\end{aligned}
\end{equation}
where $z$ is the mean in- and out-degree derived from
\begin{equation}\label{eq:generating-fuction-mean-out-degree}
    z=\frac{\partial \mathcal{G}(x,y)}{\partial x}\biggl \vert_{x=y=1} =  \frac{\partial \mathcal{G}(x,y)}{\partial y}\biggl \vert_{x=y=1}.
\end{equation}

Note that the in- and out-excess degree distribution generating functions we just derived ($\mathcal{G}^{in}_1(x)$ and $\mathcal{G}^{out}_1(x)$) refer to excess in- and out-degree distribution of a random event in the event graph.

\section{Details of analytical derivation of critical exponents}
To study properties of the event graph, we approximate it by a random directed graph with the same in- and out-degree distribution. The following sections are all based on this assumption. The validity of this assumption and the following results can be verified explicitly by empirically constructing temporal networks of different topologies and temporal dynamics and measuring scaling of quantities such as $\rho(t)$, $P(t)$, $M$, $V$ or $\rho_\text{stat}(\tau)$ \cite{longpaper}.

\subsection{Control Parameter \texorpdfstring{$\tau$}{tau}}

The mean-field rate equation for occupation density in homogeneous occupation initial condition can be constructed as 
\begin{equation}\label{eq:rate-equation}
    \partial_t \rho(t)= [Q_{out}(2) - Q_{out}(0)]\rho(t) - [Q_{out}(1) + 2Q_{out}(2)]\rho^2(t)
\end{equation}
where $Q_{out}(i) = \frac{\partial^i}{i! \partial y^i} G^{out}_1(y)$ is the excess out-degree distribution of events in the event graph. Using excess degree distribution captures the fact that in the random temporal model we are using, in- and out-degrees of events in the event graph are correlated and both are a function of degree of the event's constituting nodes in the static base network. By defining $\tau = Q_{out}(2) - Q_{out}(0)$ and $g = Q_{out}(1) + 2Q_{out}(2)$, Eq.~\ref{eq:rate-equation} turns into $\partial_t \rho(t) = \tau \rho(t) - g \rho^2(t)$ with stationary solutions at $\rho = 0$, which represents the absorbing phase, and $\tau = 0$, which corresponds to the mean-field critical point.

An event graph can be presented, without any change in reachability of any event, so that no event has an in- or out-degree larger than two (as discussed in the beginning of this section) $\tau$ and $g$ can be written as $\tau = \left \langle  Q_{out}\right \rangle  - 1$ and $g = \left \langle Q_{out}\right \rangle $.

The phase transition at $\tau = 0$ also complies with the previously know result of phase transition in random directed graphs with arbitrary degree distribution at $\frac{\partial}{\partial y} G^{out}_1(y)|_{y=1} = 1$ \cite{newman2001random}.

\subsection{Density scaling exponents \texorpdfstring{$\alpha = \beta = 1$}{alpha = beta = 1}}
For large $t$, Eq.~\ref{eq:rate-equation} has one solution for active and absorbing phases and the critical threshold $\tau = 0$
\begin{equation}
    \rho(t) = \left \{
    \begin{aligned}
        &-\tau\left(g-\frac{\tau}{\rho_0}\right)^{-1} \mathrm{e}^{\tau t}, &&\text{if}\ \tau < 0\\
        &\left(\rho_0^{-1}+gt\right)^{-1}, &&\text{if}\ \tau = 0\\
        &\frac{\tau}{g}+\frac{\tau}{g^2}\left(g-\frac{\tau}{\rho_0}\right) \mathrm{e}^{-\tau t}, &&\text{if}\ \tau > 0
    \end{aligned}
    \right.
\end{equation}
where as $t$ grows, $\rho$ approaches zero for $\tau \leq 0$ and $\rho \rightarrow \tau/g$ for the $\tau > 0$, i.e.~ asymptotically
\begin{equation}
    \rho(t) \propto t^{-1}, \text{if}\ \tau = 0
\end{equation}
and
\begin{equation}
    \rho_{stat}(\tau) \propto \tau^1, \text{if}\ \tau > 0.
\end{equation} which leads to 
\begin{equation}\label{eq:alpha-equals-beta-equals-one}
    \alpha = \beta = 1.
\end{equation}

\subsection{Rapidity-reversal symmetry \texorpdfstring{$\beta' = \beta$}{beta = beta'}}
The fact that survival of a component is measured using out-component of events in the event graph while occupation density is calculated by measuring in-component of all possibly infected nodes, hints at a symmetry in the system under time reversal. Consider $\delta t$-constrained event graph representation of temporal network $T(\mathcal{V}, \mathcal{E})$ and two sets events in bands of time $\delta t$ units of time wide, namely $E_0 = \{ e \in \mathcal{E} \mid 0 \leq t_e < \delta t \}$ and $E_t = \{ e \in \mathcal{E} \mid t  \leq t_e < t + \delta t \}$ where $t_e$ is time of activation of event $e$. Assuming $S_t \subseteq E_t$ where each member of $S_t$ appears in the out-component of at least one of the members of $E_0$ and $S_0 \subseteq E_0$ where each member of $S_0$ appears in the in-component of at least one of the members of $E_t$ (which is to say, one of the members of $S_t$). Probability of survival at time $t$ can be estimates as the fraction of nodes in $E_0$ that can reach at least a node in $E_t$, $P(t) \approx |S_0|/|E_0|$. Similarly, since in the homogeneous fully occupied case all the events in $E_0$ are occupied, the occupation density at time $t$ can be estimated as $\rho(t) \approx |S_t|/|E_t|$.

Under reversal of time $t_e \rightarrow (t + \delta t) - t_e$ the direction of the links in the event graph will revert which in turn causes switching of in- and out-component set of each node. In this scenario, occupation density is estimated by $\rho(t) \approx |S_0|/|E_0|$ which is the same as probability of survival in the original case. Conversely, probability of survival is estimated by $P(t) \approx |S_t|/|E_t|$ which is the same as occupation density in the original case. If the time-reverted event graph has the same likelihood as the original event graph, e.g.~if $\mathcal{G}^{out}_0 = \mathcal{G}^{in}_0$, this leads to the identity
\begin{equation}
    P(t) = \rho(t),
\end{equation}
which in turn, for models belonging to the DP class, leads to the celebrated rapidity-reversal symmetry:
\begin{equation}\label{eq:beta-equals-betaprime}
    \beta = \beta'.
\end{equation}

\subsection{Mean component mass exponent \texorpdfstring{$\gamma = 1$}{gamma = 1}}

The generating function for distribution of out-component sizes $H_0(y)$ is the solution to the system
\begin{equation}\label{eq:out-component-generation-function}
\begin{aligned}
    H_0(y) &= y\mathcal{G}^{out}_0(H_1(y))\\
    H_1(y) &= y\mathcal{G}^{out}_1(H_1(y))
\end{aligned}
\end{equation}
and mean out-component size can be calculated as
\begin{equation}\label{eq:mean-out-component-size}
    M = \frac{\partial H_0(y)}{\partial y}\biggl\vert_{y=1} =  1 + \mathcal{G}^{'\text{out}}_0(1) H'_1(1)\,.
\end{equation}
For $tau < 0$ where $H_1(1) = 1$ this results in a solution in form of
\begin{equation}\label{eq:m_vs_h1}
\begin{aligned}
&H'_1(1) = 1 +  \mathcal{G}^{'\text{out}}_1(1) H'_1(1) = (1 - \mathcal{G}^{'\text{out}}_1(1))^{-1} \\
&\rightarrow M = 1 + \mathcal{G}^{'out}_0(1) (1 - \mathcal{G}^{'\text{out}}_1(1))^{-1} \,.
\end{aligned}
\end{equation}

Keeping in mind the definition of control parameter $\tau = \left \langle  Q_{out}\right \rangle  - 1 =  \mathcal{G}^{'\text{out}}_1(1) - 1$ and that $\frac{\partial}{\partial y}\mathcal{G}^{\text{out}}_0( y)|_{y=1} = z$ (see Eq.~\ref{eq:generating-fuction-mean-out-degree}), we can re-write $M$ as
\begin{equation}
    M = 1 + z (-\tau)^{-1} = \frac{z - \tau}{-\tau} \,,
\end{equation}

For the special case of random $k$-regular networks we can prove that $z -\tau = 1$ which give the result $M=(-\tau)^{-1}$. More generally, to find exponent of a power-law asymptote of the form $(-\tau)^{-\gamma}$ as $\tau \rightarrow 0^-$ for any random graph we can find the solution to
\begin{equation}
\begin{aligned}
    -\gamma &= \lim_{\tau \rightarrow 0^-} \frac{\ln M}{\ln -\tau} = \lim_{\tau \rightarrow 0^-} \frac{\ln (z- \tau) - \ln -\tau}{\ln -\tau} \\
    &= \lim_{\tau \rightarrow 0^-} \frac{\ln (z- \tau)}{-\tau} -1 = -1 \, \text{if}\ 0 < z < \infty\,.
\end{aligned}
\end{equation}
Under the condition that $0 < z < \infty$, the limit term is equal to zero, resulting in $\gamma = 1$. Given the fact that, assuming probability of co-occurrence of adjacent events is negligible, the maximum out-degree of the event graph is 2, if the mean out-degree is above zero at $\tau = 0$, a power relation with exponent $\gamma = 1$ estimates mean component mass. 

\section{Alternative analogue for mean cluster mass}
\begin{figure*}
    \centering
    \includegraphics[width=1\linewidth]{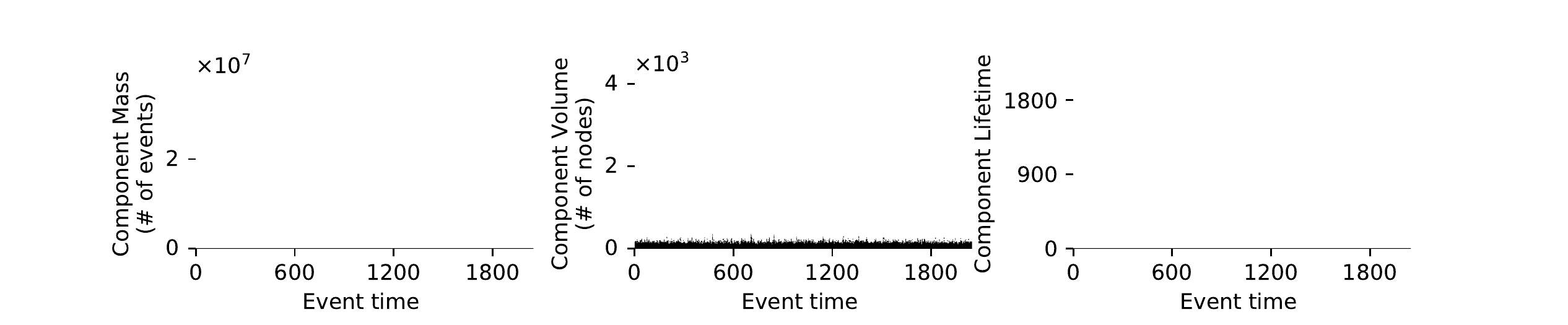}
    \includegraphics[width=1\linewidth]{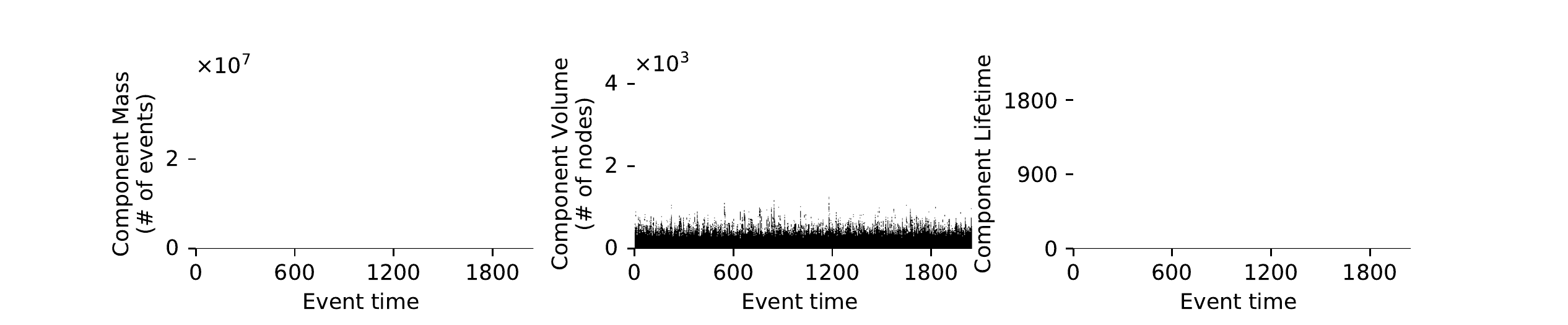}
    \includegraphics[width=1\linewidth]{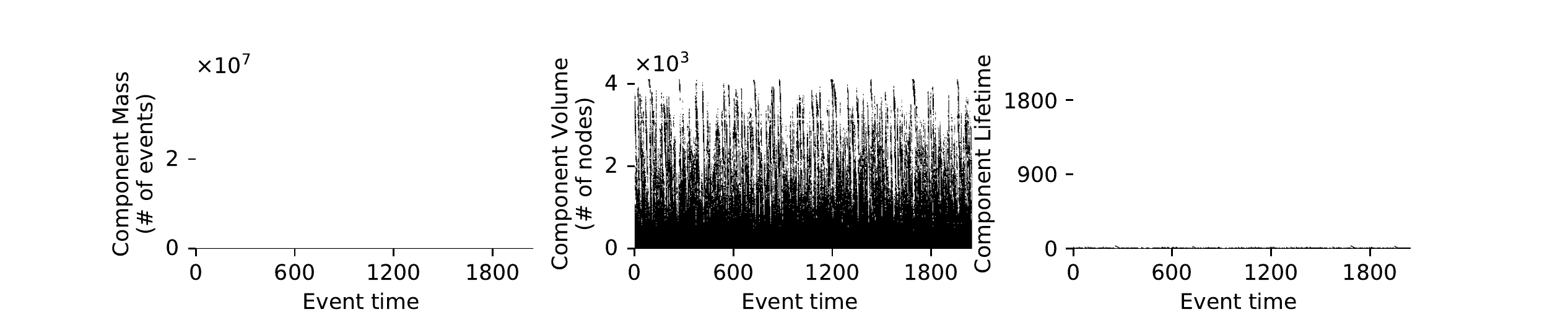}
    \includegraphics[width=1\linewidth]{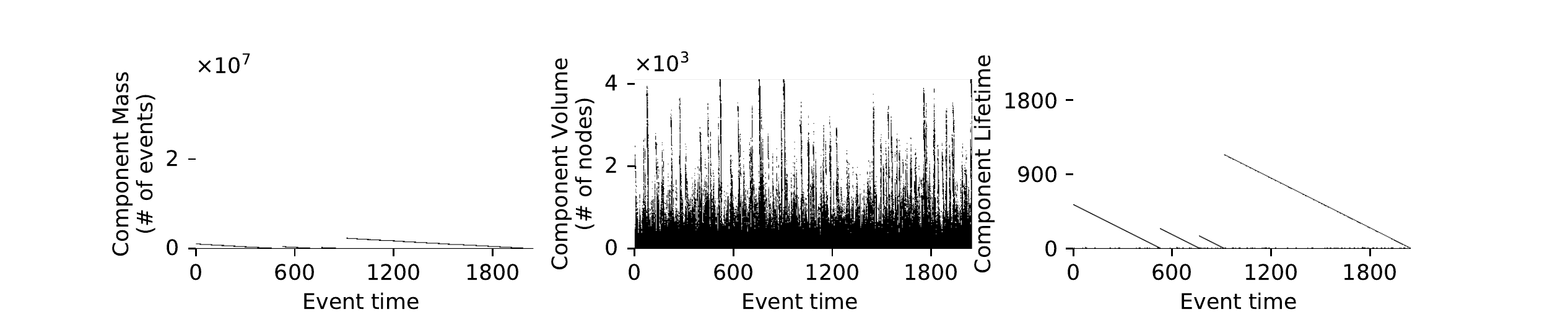}
    \includegraphics[width=1\linewidth]{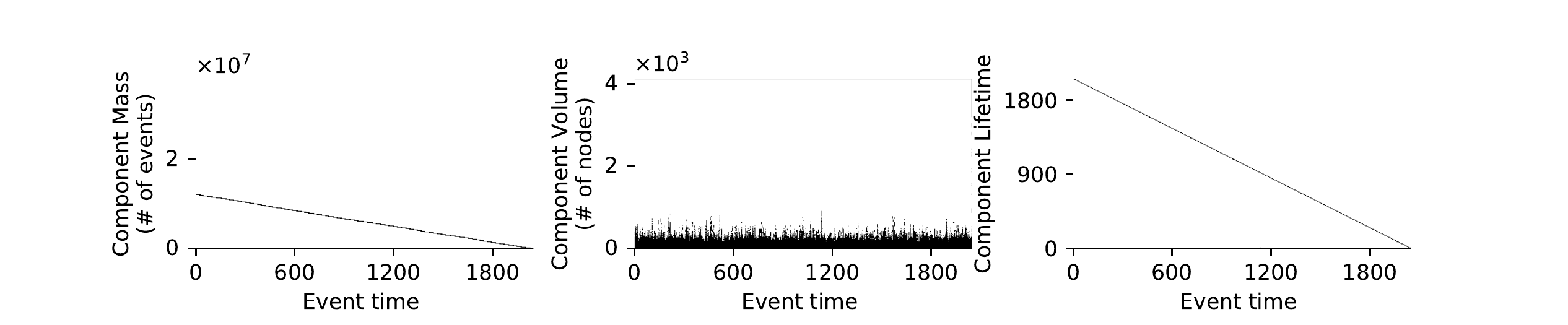}
    \caption{Out-component size estimates of all events of a 9-regular network with Poisson process link activations $\lambda=1$ for (a) $\delta t=0.07$, (b) $\delta t=0.08$, (c) $\delta t=0.08802$, (d) $\delta t=0.092$ and (e) $\delta t=0.1$.}
    \label{fig:regular-out-component-sizes}
\end{figure*}

In classic DP, mean cluster mass $M$ is defined as the integration of the pair-connectedness function across all nodes and time. Based on how we established parallels between spatial dimensions and the base networks and a the possibility of defining pair-connectedness function as existence of a $\delta t$-limited time-respecting path between a pair of nodes at different times, this might translates more directly to the sum of time-length of all infections started from a random event, or in other words total duration of sickness for all people. This would also imply that the spreading processes start at random nodes and times, as opposed to starting at random events as we use in the manuscript. Values for mean cluster mass $M$, as well as mean cluster volume $V$ and mean cluster lifetime $T$ for random 9-regular networks with Poisson lik activations are plotted in Fig.~\ref{fig:regular-out-component-sizes} for different values of $\delta_t$.

\section{Description of the real-world temporal network data sets}
Four real-world temporal networks were used for demonstrating the measurement of the quantities and phase transition. These are the same datasets used previously for developing the algorithmic method which form the backbone of the more empirical parts of the current manuscript \cite{badiemodiri2020efficient}.

The air transportation network dataset is a recording of 180\,112 flights between 279 airports in the United States, gathered from the website of the The Department of Transportation's Bureau of Transportation Statistics website in 2017 \cite{bts2017air}. The public transportation network is the set of all 664\,138 trips during a typical Monday in Helsinki in 2018, where a trip is one public transportation vehicle moving from one of the 6\,858 bus, metro and ferry stations to the next \cite{kujala2018collection}. The twitter dataset is a set of 266\,179\,671 mentions (counting replies) of 17\,313\,552 user handles \cite{yang2011patterns}. Finally, the mobile phone call dataset is set of 324\,576\,400 phone calls between over 5\,193\,086 mobile phone subscribers \cite{karsai2011small}. A few thousand events were removed from the beginning of the twitter dataset to eliminate a weeks-wide gap in the gathered data.

The first two networks, air and public transportation networks, were processed as directed, delayed temporal networks where each event has a duration as well as a starting time $e = (v_1, v_2, t, d)$ where two events are adjacent if the second event starts after the duration of the event is finished and the tail node of the second event is the same as the head node of the first event, e.g.~the first plane lands in the destination airport before the second one takes off from that airport. The waiting time $\delta t$ then refers to the time between end of the first event to the beginning of the second one.

There is an argument for measuring waiting time in delayed temporal networks from the beginning of the first event for some processes such as disease spreading. For example, a disease that gets healed less than an hour after infecting someone has a very low chance of spreading through air travel where most trips take longer than that. That method was not used in this manuscript. The second pair of networks, twitter and mobile networks, were treated like undirected, instantaneous temporal networks.

The real-world networks show high degrees of temporal heterogeneity, daily/weekly patterns, peaks at spacial hours of the day or at special days of the year or local or global outages. Measuring representative values for static density $\rho_\text{stat}$ and susceptibility $\chi$ for real-world networks would need special consideration. Our current method for measuring these quantities in the homogeneous, fully-occupied initial condition is dependent on the level of activity of the initial time $t_0$ of the dataset as well as existence of unlikely periods of very low or very high activity as a result of natural disasters, real-world happenings or simply failure of the measurement apparatus or the measured system. To average out any such outliers we split the original data into 64 equal time windows of time $T/2$ (for air and public transport) and $T/16$ (for mobile and twitter) where $T$ is the time window of the original dataset, each starting at a random point in time.

\begin{figure*}[h]
    \centering
    \includegraphics[width=1\linewidth]{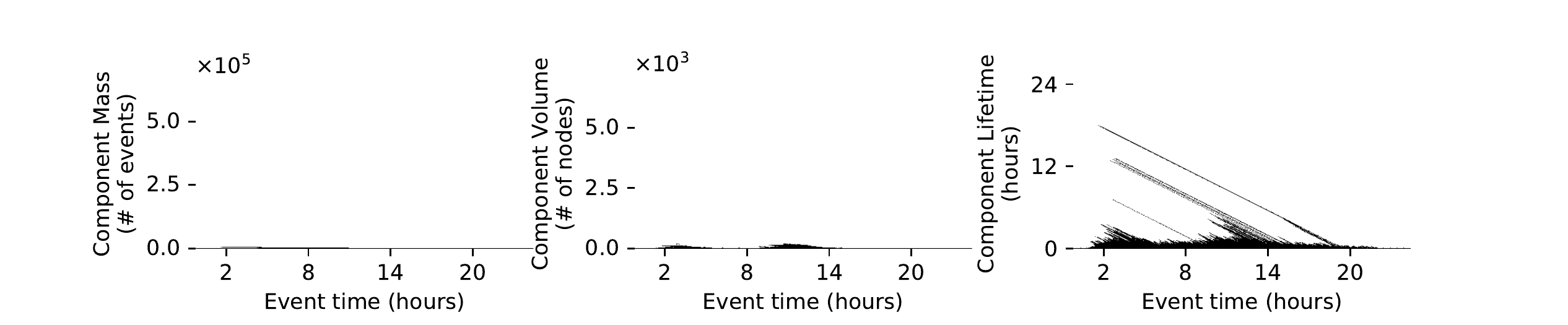}
    \includegraphics[width=1\linewidth]{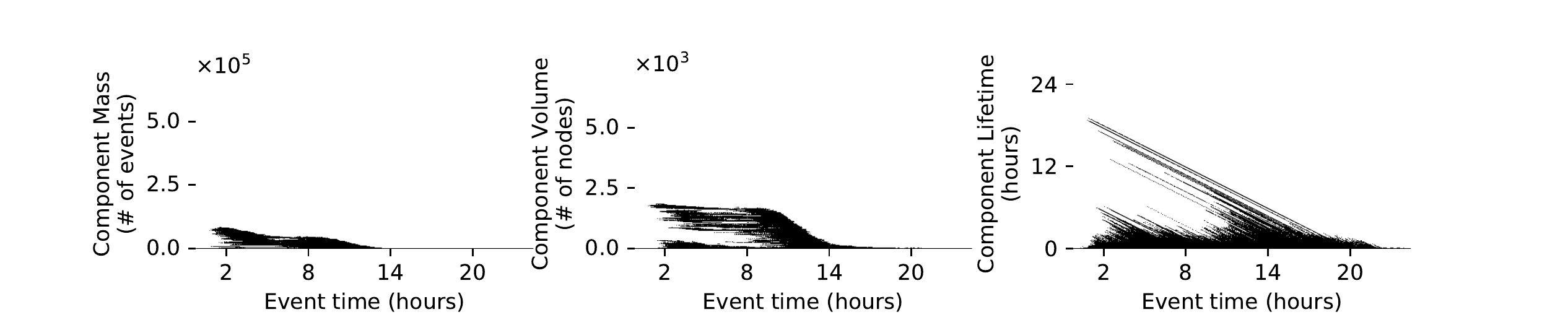}
    \includegraphics[width=1\linewidth]{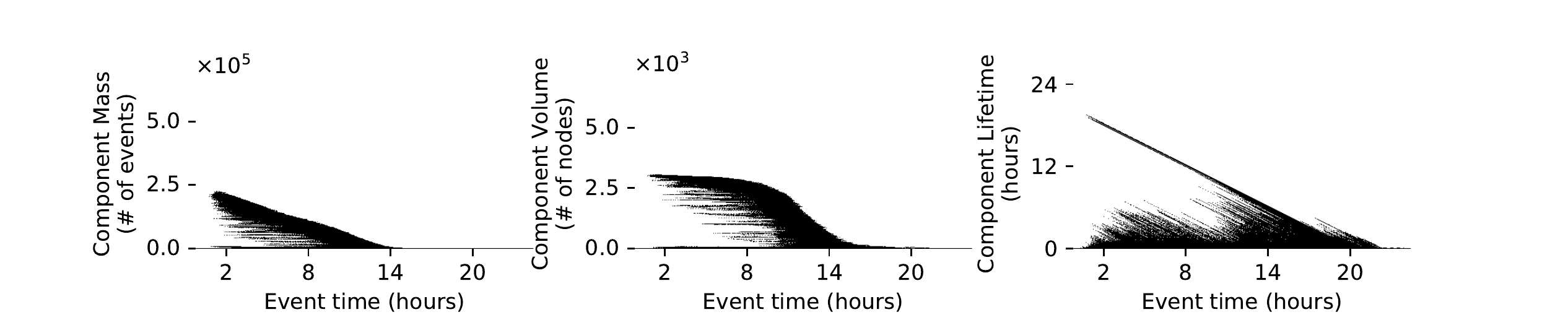}
    \includegraphics[width=1\linewidth]{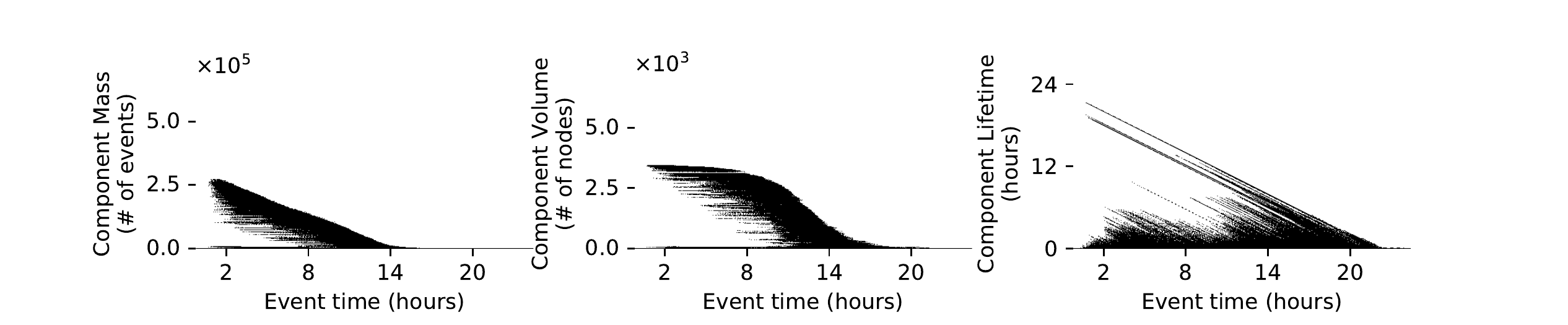}
    \includegraphics[width=1\linewidth]{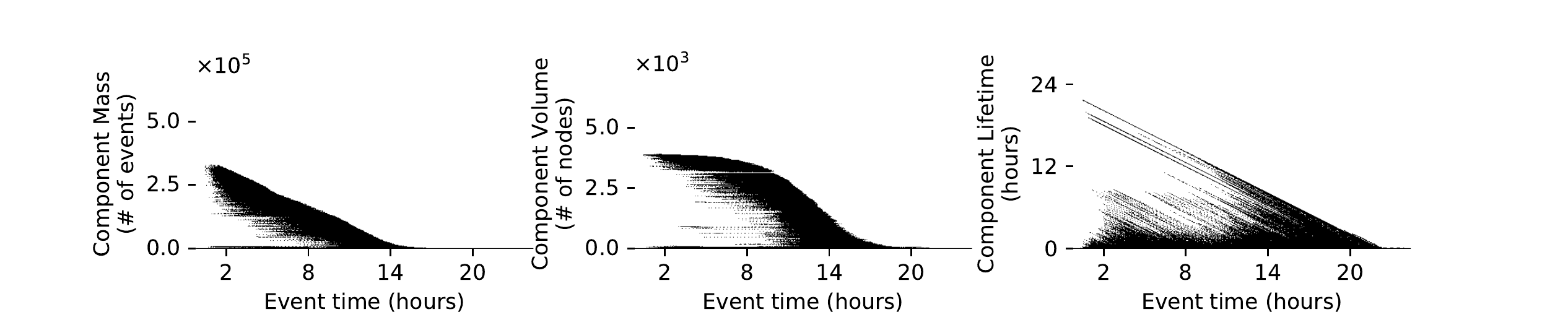}
    \caption{Helsinki public transport network out-component size estimates for (a) $\delta t=300$, (b) $\delta t=500$, (c) $\delta t=\delta t_c=670$, (d) $\delta t=800$ and (e) $\delta t=1000$ seconds.}
    \label{fig:transport-out-component-sizes}
\end{figure*}

\begin{figure*}[h]
    \centering
    \includegraphics[width=1\linewidth]{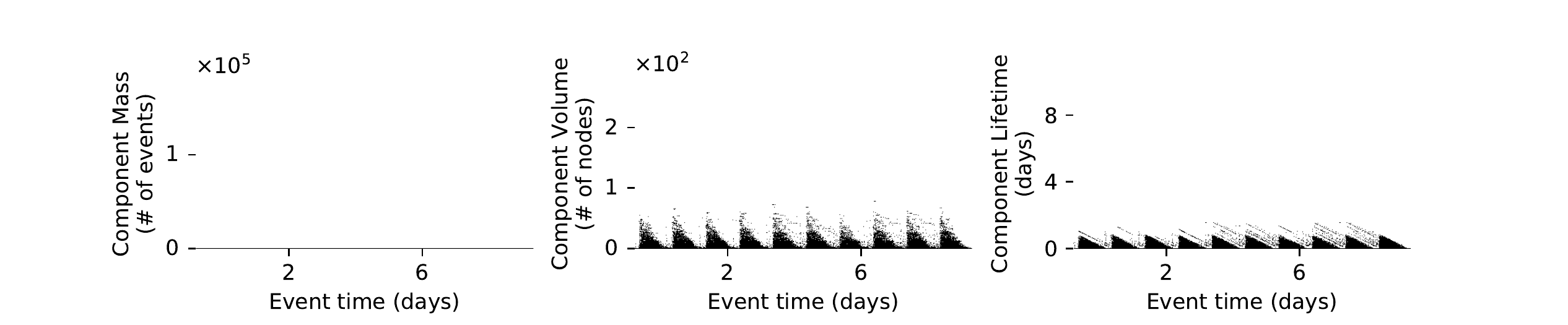}
    \includegraphics[width=1\linewidth]{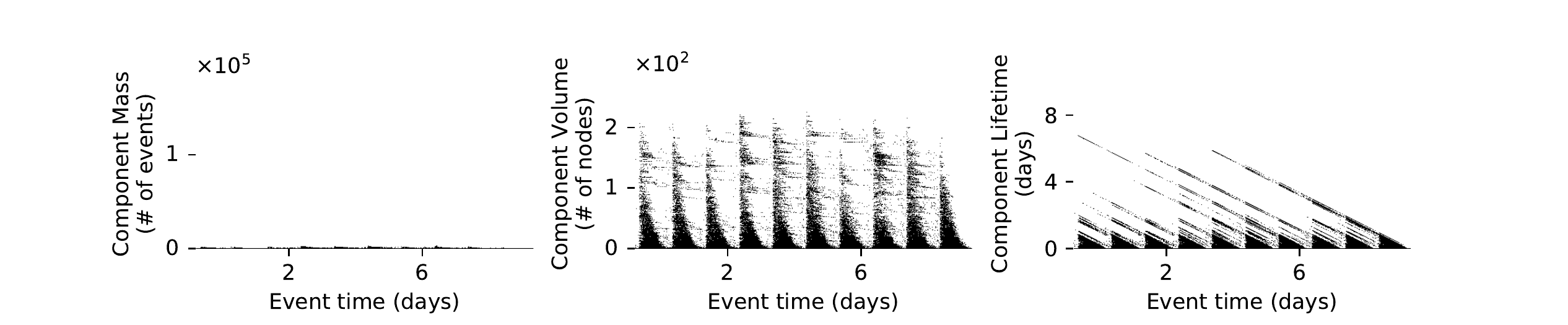}
    \includegraphics[width=1\linewidth]{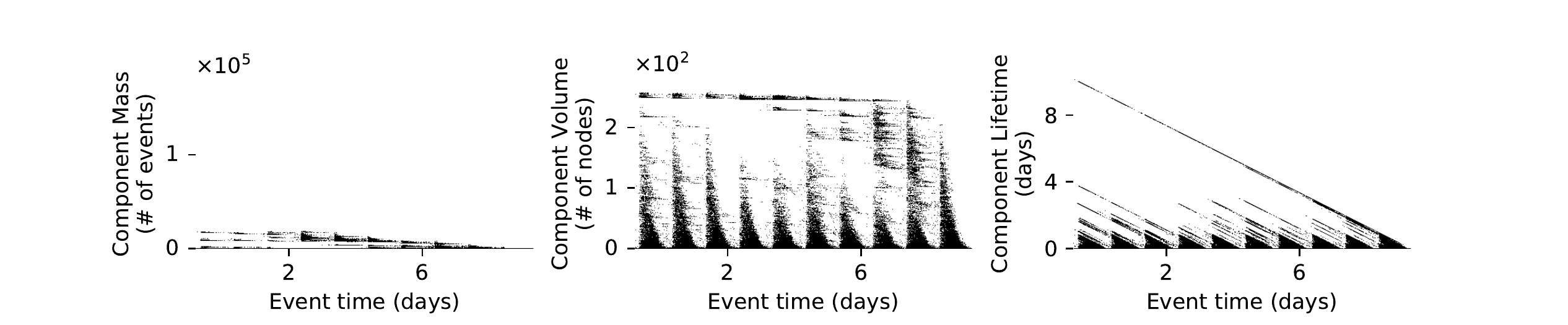}
    \includegraphics[width=1\linewidth]{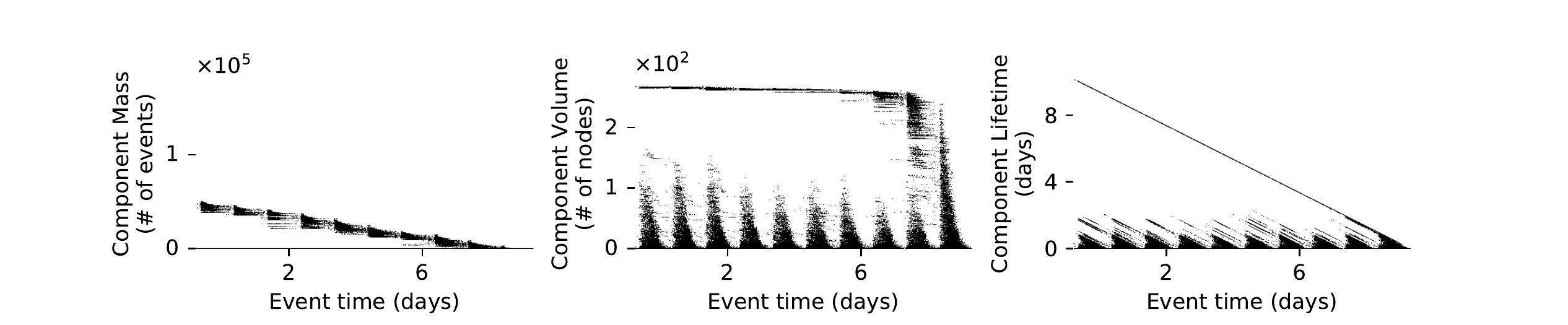}
    \includegraphics[width=1\linewidth]{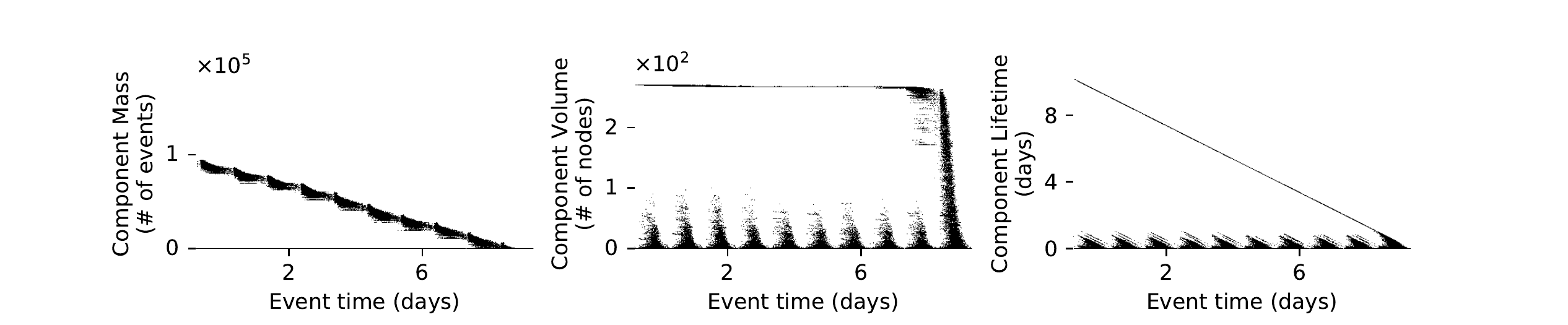}
    \caption{Air transport network out-component size estimates for (a) $\delta t=300$, (b) $\delta t=400$, (c) $\delta t=\delta t_c=470$, (d) $\delta t=600$ and (e) $\delta t=800$ seconds.}
    \label{fig:air-out-component-sizes}
\end{figure*}

\begin{figure*}[h]
    \centering
    \includegraphics[width=1\linewidth]{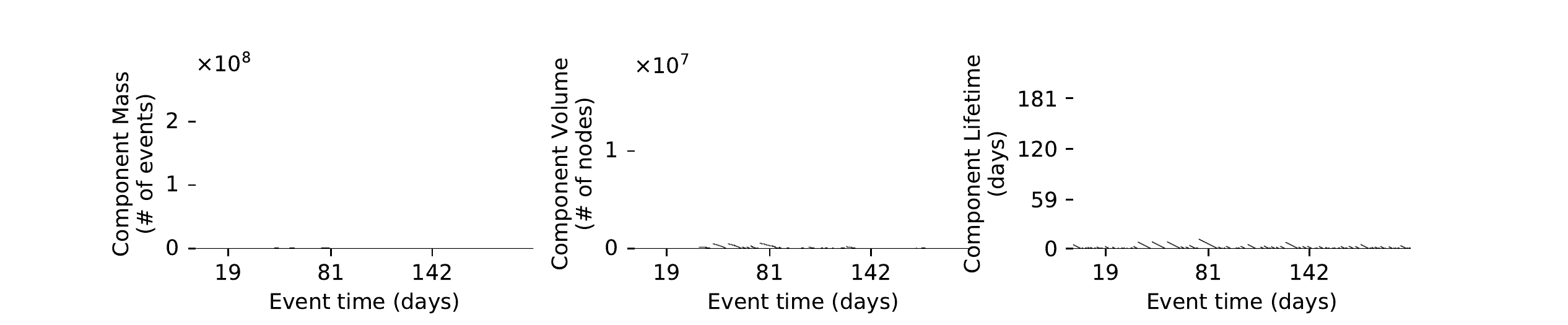}
    \includegraphics[width=1\linewidth]{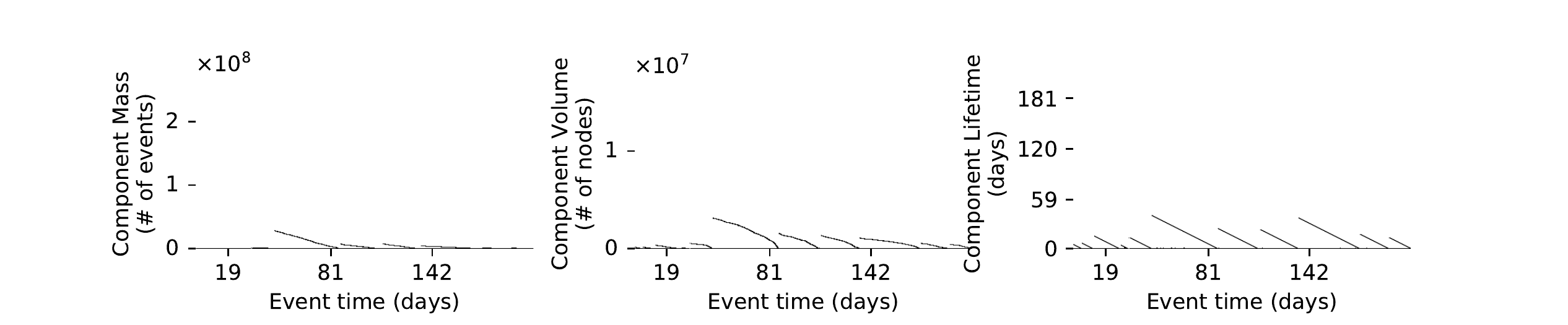}
    \includegraphics[width=1\linewidth]{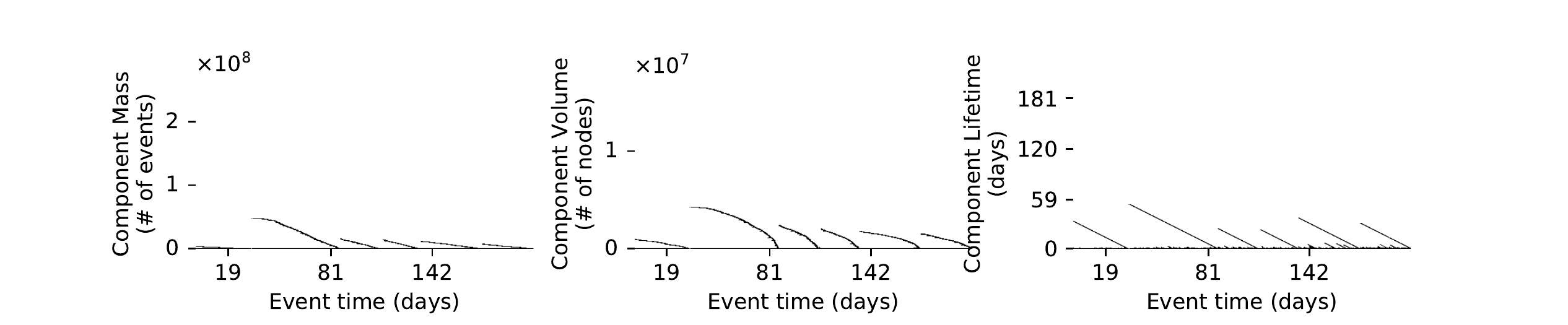}
    \includegraphics[width=1\linewidth]{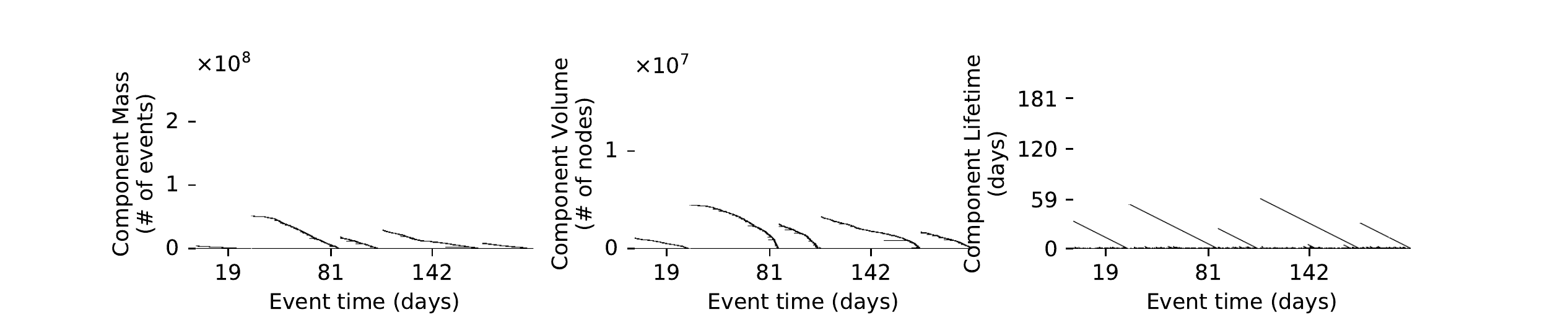}
    \includegraphics[width=1\linewidth]{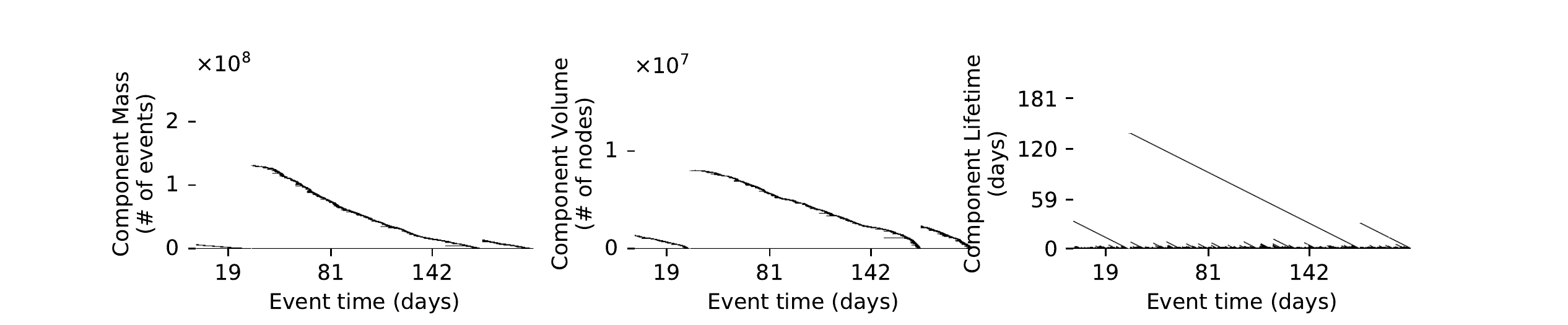}
    \caption{Twitter mention network out-component size estimates for (a) $\delta t=200$, (b) $\delta t=1200$, (c) $\delta t=3600$, (d) $\delta t=\delta t_c=4800$ and (e) $\delta t=12000$ seconds.}
    \label{fig:twitter-out-component-sizes}
\end{figure*}

\begin{figure*}[h]
    \centering
    \includegraphics[width=1\linewidth]{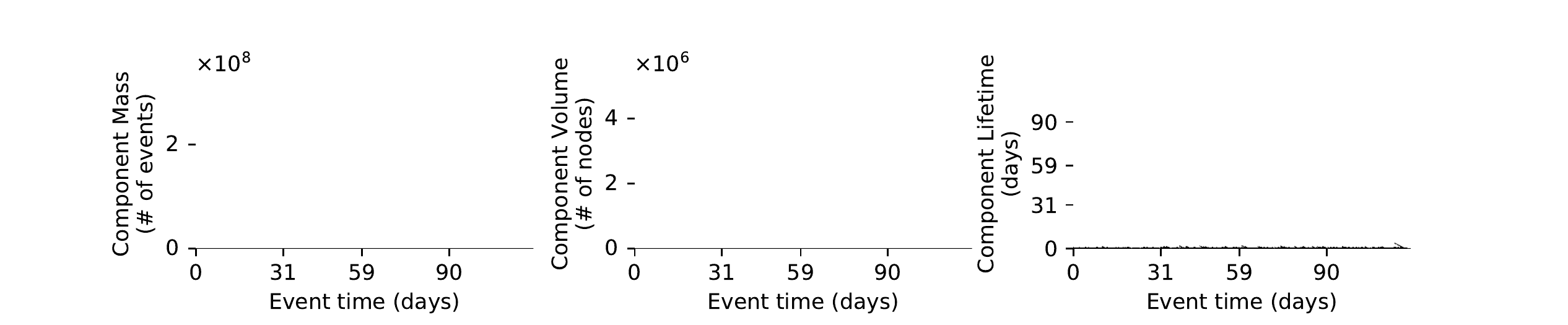}
    \includegraphics[width=1\linewidth]{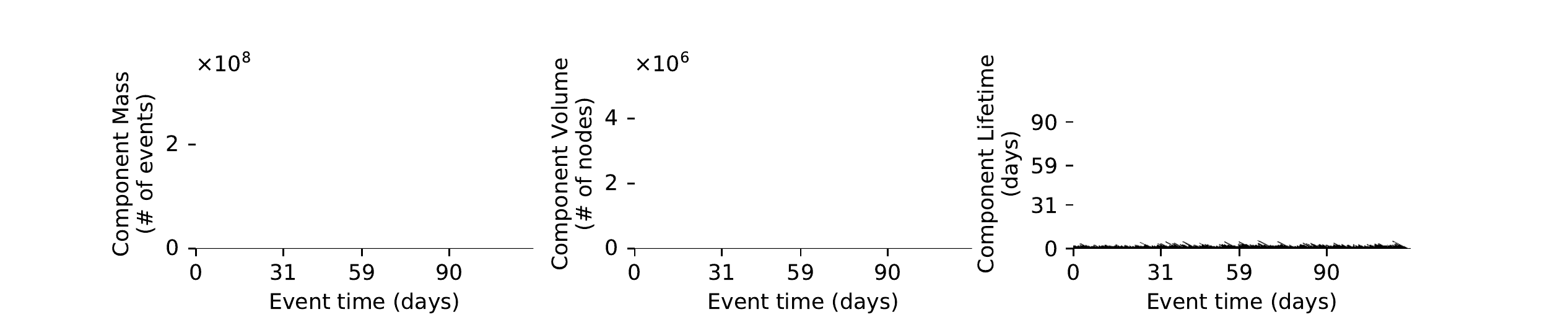}
    \includegraphics[width=1\linewidth]{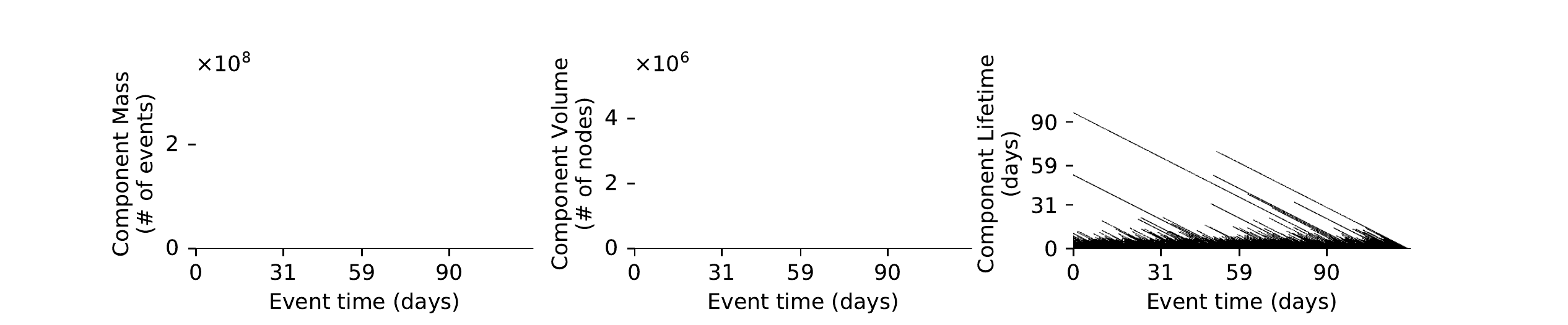}
    \includegraphics[width=1\linewidth]{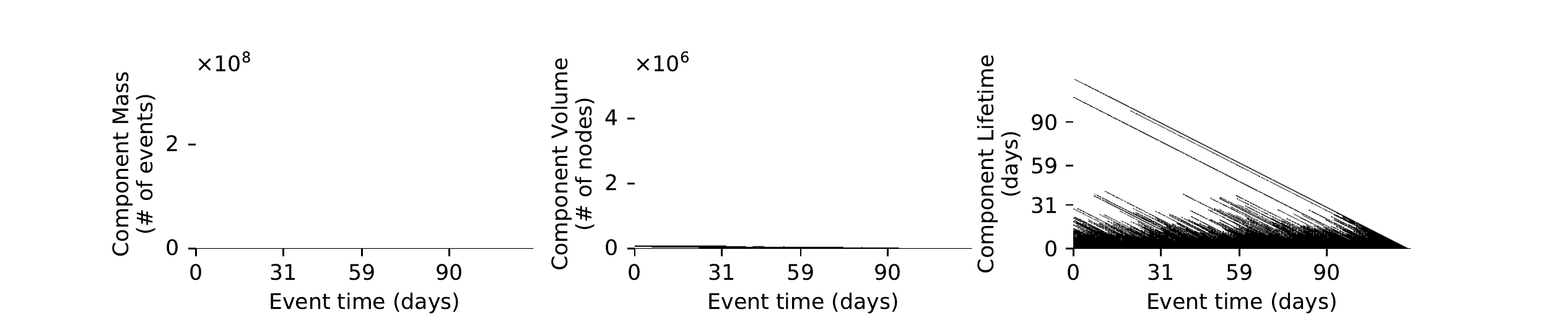}
    \includegraphics[width=1\linewidth]{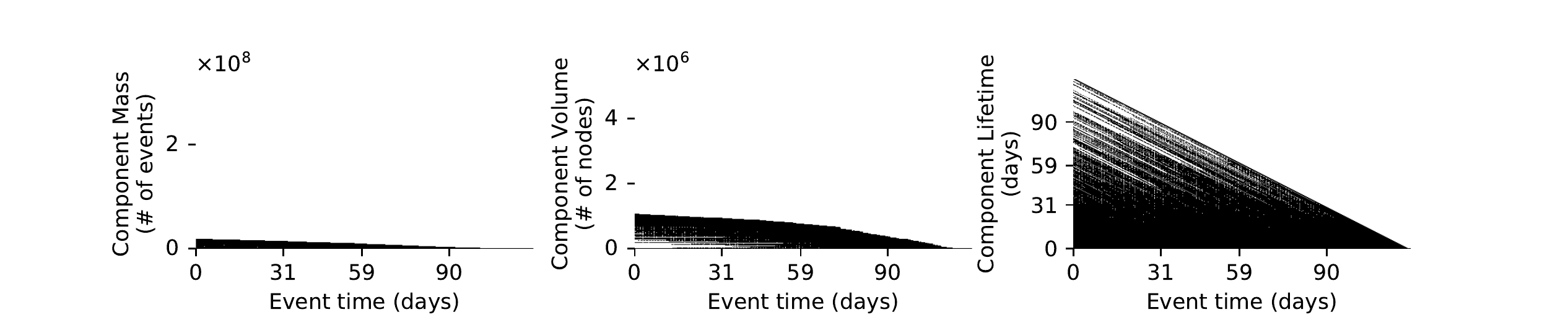}
    \caption{Mobile call network out-component size estimates for (a) $\delta t=2$, (b) $\delta t=4$, (c) $\delta t=\delta t_c=7.5$, (d) $\delta t=9$ and (e) $\delta t=14$ hours.}
    \label{fig:mobile-out-component-sizes}
\end{figure*}

Distribution of mass, volume and lifetime for each event in the event graph can be seen Figs.~\ref{fig:transport-out-component-sizes}, \ref{fig:air-out-component-sizes}, \ref{fig:twitter-out-component-sizes} and \ref{fig:mobile-out-component-sizes} for Helsinki public transport, Air transport, Twitter and Mobile datasets respectively.

\section{Joint degree distribution in real-world temporal network event graphs}
Symmetry of the joint in-~and out-degree distribution of the event graph ($\forall_{i,o} \, p^{in,out}_{i,o} = p^{in,out}_{o,i}$) was discussed as a condition for $\beta = \beta'$ under a mean-field assumption of connectivity. While real-world networks often have correlations and inhomogeneities that affect connectivity, it is still interesting to verify the validity of this condition in real-world networks.

Observation of joint in- and out-degree of 100\,000 random events of the event graphs for the Mobile and Twitter networks, the two largest real-world systems studied, show degree distributions (degrees 0 to 2) estimated at: 
\begin{equation}
p^{in,out} =
\begin{pmatrix}
0.11063 & 0.16668 & 0.02007\\
0.16263 & 0.37322 & 0.05916\\
0.02028 & 0.05570 & 0.02359
\end{pmatrix}
\end{equation}
and
\begin{equation}
p^{in,out} =
\begin{pmatrix}
0.30426 & 0.10793 & 0.02340\\
0.11109 & 0.20426 & 0.05220\\
0.02082 & 0.04992 & 0.03393
\end{pmatrix}
\end{equation}
respectively at $\delta t = 7.5\ \text{hours}$ and $\delta t = 25\ \text{minutes}$.
\bibliography{citations}